\documentclass[aps,prb,twocolumn,letterpaper,superscriptaddress]{revtex4-2}

\usepackage{amsmath}
\usepackage{amssymb}
\usepackage{amsfonts}
\usepackage{bm}
\usepackage{mathrsfs}
\usepackage{tensor}
\usepackage{dsfont}
\usepackage{esint}
\usepackage{comment}

\usepackage[usenames,dvipsnames]{xcolor}
\usepackage[hyperindex,pdftex, breaklinks,colorlinks = true,linkcolor = blue,urlcolor=blue,citecolor=blue]{hyperref}

\usepackage{physics}
\usepackage[section]{placeins}

\begin{document}
	
	\title{Chern insulators in two and three dimensions: A global perspective}

	\author{Jason G. Kattan}
	\email{jkattan@physics.utoronto.ca}
	\affiliation{Department of Physics, University of Toronto, Toronto, Ontario M5S 1A7, Canada}
 
	\author{J. E. Sipe}
	\email{sipe@physics.utoronto.ca}
	\affiliation{Department of Physics, University of Toronto, Toronto, Ontario M5S 1A7, Canada}
	
	\date{\today}
	
	\begin{abstract}
             We introduce a second-quantized field theory for Chern insulators in which the Hamiltonian features a static vector potential that has the periodicity of the crystal's lattice and spontaneously breaks time-reversal symmetry in the system's ground state. Such a vector potential generates a magnetic field at the microscopic level that may be thought of as arising from local moments associated with one or more magnetic ions in each unit cell. Considering spinor electrons, we study the Chern invariants characterizing the topology of the occupied valence bands of Chern insulators in both two and three dimensions -- the Chern number and the Chern vector, respectively -- and we derive novel expressions for these topological invariants that are globally defined across the Brillouin zone and involve the full band structure of the system. We also study the long-wavelength response of a Chern insulator to electromagnetic fields at finite frequency, generalizing the quantum anomalous Hall effect in the static limit to the optical regime. 
        \end{abstract}
	
	\maketitle

%%%%%%%%%%%%%%%%%%%%%%%%%%%%%%%%%%%%%%%%%%

\section{Introduction}\label{Sec:Introduction}

``Quantum materials" are special phases of condensed matter for which basic quantum mechanical descriptions, often involving semiclassical particles, are insufficient to fully characterize the underlying physics \cite{QuantMatARPES}. Particularly interesting examples are \textit{topological} quantum materials, where even a detailed knowledge of the spectral properties of a system's band structure is insufficient, and additional topological invariants must be specified \cite{Bernevig}. There are several different kinds of such materials, which are grouped together into various classes in the ``tenfold way" classification of topological insulators and superconductors \cite{Kitaev2009,Ryu2010,Freed}. These groupings are distinguished by which combination of time-reversal symmetry, particle-hole symmetry, and chiral (or ``sublattice") symmetry is respected by the system's Hamiltonian \cite{Ludwig2016}. But there is one class for which \textit{none} of these symmetries hold, and thus the topological properties of these ``Class A" insulators are not protected by any of these symmetries. In three or fewer dimensions, the only such Class A topological materials are \textit{Chern insulators} \cite{Kaufmann}.

Chern insulators are band insulators with a periodic lattice structure and spontaneously broken time-reversal symmetry, where the occupied ``valence" bands are topologically nontrivial, being are described by one or more integer-valued invariants called ``Chern numbers" \cite{VanderbiltBook, Resta2011, Monaco2017, Troyer2016}. Chern numbers are \textit{topological} invariants in the sense that they are unchanged under continuous and adiabatic variations of the system's Hamiltonian, and they arise, for example, in the description of the quantum anomalous Hall effect when a static electric field is applied \cite{QAHcolloquium}. Two-dimensional Chern insulators are characterized by one such Chern number, which has been referred to as a ``strong" topological invariant because of its robustness against impurities and lattice dislocations \cite{Ryu2010}. Meanwhile, there are a triple of Chern numbers characterizing a Chern insulator in three dimensions, which are ``weak" topological invariants in the sense that any disorder in the lattice will generally negate their topological quantization. However, they can be assembled into a vector-valued invariant, called the Chern vector, that is, in fact, robust against disorder in the lattice \cite{WeakStrongTI}.

There have been a number of materials found to support a Chern insulator phase, the first examples of which were thin films of magnetically doped $\mathrm{Bi}_2\mathrm{Te}_3$ and $\mathrm{Sb}_2\mathrm{Te}_3$ \cite{QAHExperiment2, QAHExperiment3}, and three-dimensional structures formed from stacking layers thereof \cite{QAHExperiment5}. Magnetic doping, such as by the inclusion of one or more chromium ($\mathrm{Cr}$) or vanadium ($\mathrm{V}$) atoms in each unit cell, leads to the breaking of time-reversal symmetry in these materials. It has also been demonstrated experimentally \cite{QAHExperiment6} that thin films of the intrinsic magnetic material $\mathrm{Mn}\mathrm{Bi}_2\mathrm{Te}_4$ become Chern insulators at low temperatures for an odd number of septuple layers, where time-reversal is broken by the monoatomic lattice of manganese ($\mathrm{Mn}$) atoms in each septuple layer. In all of these materials, a static magnetic field is generated at the microscopic level by the intrinisic magnetic moments of some transition-metal atoms in the crystal. Chern insulator phases have also been realized in some Moiré materials formed from graphene \cite{QAHExperiment8} and transition-metal dichalcogenides \cite{QAHExperiment7}, although in these twisted multilayer systems the breaking of time-reversal symmetry occurs due to strong interlayer electronic correlations, instead of long-range magnetic order \cite{QAHcolloquium}.

One of the first theoretical models for Chern insulators was the tight-binding model on a honeycomb lattice introduced by Haldane \cite{Haldane}, where time-reversal symmetry is broken by the inclusion of a complex phase multiplying the next-to-nearest-neighbor hopping terms in the Hamiltonian. This complex phase was later interpreted as a kind of Peierls phase due to the presence of a cell-periodic vector potential intrinsic to the system \cite{Sticlet2012}. Extensions of this model were further studied within the ``modern theories of polarization and magnetization" \cite{KingSmith1993, Resta1994,Resta2010}, where macroscopic notions of polarization and magnetization in a Chern insulator can be defined \cite{Niu2007, Vanderbilt2009,Resta2006}. Meanwhile, more recent studies of Chern insulators generally follow one of two approaches. In the first approach, tight-binding models are employed by means of effective Hamiltonians with parameters chosen to guarantee that the system is in a Chern insulator phase \cite{Sticlet2012,LatticeCI1,LatticeCI2,LatticeCI3}. A common choice is a ``Dirac-type" Hamiltonian \cite{TseMacdonald}, motivated by the presumption that for finite Chern insulators there must be effectively \textit{linear} band crossings at the boundary of the material when there are band inversions in the bulk; such band crossings at the boundary are said to be ``chiral," and lead to the metallic surface states \cite{VanderbiltBook} that are characteristic of topological insulators. The second approach is based on modern developments in quantum geometry \cite{LiuQG}, where topological properties \cite{Mera1,Mera2} and response tensors \cite{Ghosh} are derived from the so-called ``quantum geometric tensor." 

Here we introduce a third approach. Motivated by the observation that time-reversal symmetry is spontaneously broken in experimental realizations of Chern insulators by a static magnetic field that is intrinsic to the crystal, we introduce a Hamiltonian field theory that is formulated within second-quantized quantum mechanics in which the Hamiltonian involves a static vector potential that respects the periodicity of the lattice. It generates a magnetic field at the microscopic level that itself respects the lattice periodicity \cite{CImanuscript1}, persists in the electronic ground state, and can be taken to be generated by local moments associated with one or more magnetic ions in each unit cell. We can then use basic results from condensed matter physics \cite{Girvin}, such as Bloch's theorem \cite{Bloch}, to study the electron dynamics, which are described by a second-quantized electron field operator. We focus on spin-$1/2$ electrons in the ``independent-particle approximation," where the electron-electron interactions are either neglected or treated at the mean-field level; the inclusion of electronic correlations beyond the Hartree-Fock approximation is a topic for future work.

We derive novel expressions for the topological invariants characterizing Chern insulators in two and three dimensions -- the Chern number and Chern vector of the occupied valence bands, respectively \cite{Kattan2} -- that are globally defined across the Brillouin zone and involve the full band structure of the system. These expressions, which involve such fundamental quantities as velocity matrix elements and Bloch functions, vanish in the presence of any of the three discrete symmetries that feature in the tenfold way classification scheme. This confirms that our very general Hamiltonian is appropriate to model Chern insulators, for which all three of these symmetries are necessarily broken \cite{Ryu2010}. Our expressions for these ``Chern invariants" also respect the symmetries of the crystal's lattice. Interestingly, we find that it is possible, at least in principle, to have a Chern insulator that respects inversion symmetry, unlike in time-reversal invariant topological insulators where the combination of time-reversal symmetry and inversion symmetry forces the Berry curvature to vanish. Although our ``global expressions" involve a static vector potential (through velocity matrix elements) instead of the magnetic field it generates, we show that they are still invariant under $\mathrm{U}(1)$--gauge transformations of this vector potential. In previous work \cite{CImanuscript1} we used this model to study the ground-state properties of a Chern insulator at zero temperature, where we derived macroscopic expressions for polarization and magnetization related to, but not equivalent to, those obtained from an analysis based on the ``modern theories." 

We also study the long-wavelength response of a Chern insulator to both static and finite-frequency electric fields in the linear regime, generalizing an expression for the conductivity tensor that was derived in recent work \cite{Kattan2} for spinless electrons to include the spin degree of freedom. From this conductivity tensor, which is the sum of a dynamical ``Kubo conductivity tensor" and a static term describing the quantum anomalous Hall effect, we introduce an effective dielectric tensor with which optical properties of Chern insulators can be studied. We investigate the causal properties of this dielectric tensor, which are expressed in terms of a set of generalized Kramers-Kronig relations \cite{lucarini2005kramers}, and we also demonstrate the equivalence between our conductivity tensor and that which is obtained from an analysis based on the Kubo-Greenwood formula \cite{Kubo,Greenwood}. This is an interesting result because of the difference in the formalisms used in the derivation of this tensor; our formalism, however, can be used in a straightforward way to derive corrections to this long-wavelength conductivity tensor coming from spatial dispersion effects, and even permits the derivation of nonlinear response tensors to arbitrary order in the electric and magnetic fields \cite{Mahon2020a, Duff1}.

We begin in Sec. \ref{Sec:HamiltonianTheory} with a careful formulation of this Hamiltonian field theory within second-quantized quantum mechanics, where we use standard results such as Bloch's theorem \cite{Bloch} to study the spectral properties of a Chern insulator. Then in Sec. \ref{Sec:ChernInvariants} we define the topological invariants characterizing bulk Chern insulators in both two and three dimensions, and derive global expressions for these Chern invariants in terms of the quantities introduced in Sec. \ref{Sec:HamiltonianTheory}; we study the symmetry properties of this field theory and our global expressions for the Chern invariants in Appendix \ref{Appendix:Symmetries}. And in Sec. \ref{Sec:LinearResponse} the long-wavelength response of a Chern insulator to electric and magnetic fields is discussed in detail; we show in Appendix \ref{Appendix:KuboGreenwood} that our long-wavelength conductivity tensor is in agreement with the Kubo-Greenwood formula.

%%%%%%%%%%%%%%%%%%%%%%%%%%%%%%%%%%%%%%%%%%

\section{Hamiltonian field theory}\label{Sec:HamiltonianTheory}

To begin, we consider a Chern insulator in $d$ dimensions at zero temperature, focusing on bulk crystals in two and three dimensions. A \textit{bulk} (or \textit{nominally infinite}) \textit{crystal} in $d$ dimensions is a countably infinite subset $\mathscr{C}$ of the $d$-dimensional Euclidean space $\mathbb{R}^d$ representing the locations of the ions in the crystal, which comes equipped with the additive action of a lattice $\Gamma$ of discrete translations between the same ions in different unit cells \cite{Freed}. It will be useful to fix a $d$-tuple of primitive lattice vectors spanning the crystal's lattice,
\begin{align}
    \Gamma = \mathrm{span}_{\mathbb{Z}}\big(\{\bm{a}_1,\dots,\bm{a}_d\}\big),
    \label{lattice}
\end{align}
and we refer to elements $\bm{R} \in \Gamma$ as the \textit{lattice sites}. The duality relations $\bm{a}_{\alpha}\cdot\bm{g}_{\beta} = 2\pi \delta_{\alpha\beta}$ yield an associated collection of reciprocal lattice vectors spanning the crystal's reciprocal lattice,
\begin{align}
    \Gamma^* = \mathrm{span}_{\mathbb{Z}}\big(\{\bm{g}_1,\dots,\bm{g}_d\}\big).
    \label{reciprocallattice}
\end{align}

We work in the frozen-ion approximation, in which the ions in the crystal are in a fixed configuration and do not move with respect to our chosen reference frame. Then the electrostatic potential energy $\mathrm{V}_{\Gamma}(\bm{x})$ generated by the ions in the crystal $\mathscr{C}$ is invariant under translations in the crystal's lattice, with $\mathrm{V}_{\Gamma}(\bm{x} + \bm{R}_{\mathrm{t}}) = \mathrm{V}_{\Gamma}(\bm{x})$ for all lattice translations $\bm{R}_{\mathrm{t}} \in \Gamma$. The electronic degrees of freedom are described by a pair of electron field operators $\hat{\psi}_{\sigma}(\bm{x},t)$ labelled by the $\mathrm{SU}(2)$ spinor index $\sigma \in \{\uparrow,\downarrow\}$ and satisfying the equal-time anticommutation relations
\begin{align}
    \big[\hat{\psi}_{\sigma}(\bm{x},t), \hat{\psi}_{\sigma'}^{\dagger}(\bm{y},t)\big]_+ &= \delta_{\sigma\sigma'} \delta(\bm{x}-\bm{y}),\nonumber \\
    \big[\hat{\psi}_{\sigma}(\bm{x},t), \hat{\psi}_{\sigma'}(\bm{y},t)\big]_+ &= \big[\hat{\psi}_{\sigma}^{\dagger}(\bm{x},t), \hat{\psi}_{\sigma'}^{\dagger}(\bm{y},t)\big]_+ = 0,
    \label{anticommutationrelations}
\end{align}
which guarantees that the spin-$1/2$ electrons obey the Fermi-Dirac statistics associated with the Pauli exclusion principle \cite{Girvin}. We assemble these into a \textit{$2$-spinor} electron field operator $\hat{\psi}_0(\bm{x},t) = (\hat{\psi}_{\uparrow}(\bm{x},t), \hat{\psi}_{\downarrow}(\bm{x},t))^{\mathrm{T}}$, and the dynamics of this field operator is governed by the Heisenberg equation
\begin{align}
    i\hbar \frac{\partial \hat{\psi}_0(\bm{x},t)}{\partial t} = \big[\hat{\psi}_0(\bm{x},t), \hat{H}_0(t)\big]_-,
    \label{TDSE}
\end{align}
involving the second-quantized Hamiltonian
\begin{align}
    \hat{H}_0(t) = \int d\bm{x}\, \hat{\psi}_0^{\dagger}(\bm{x},t)\mathcal{H}_0(\bm{x})\hat{\psi}_0(\bm{x},t).
    \label{H0}
\end{align}
Here we restrict ourselves to the independent-particle approximation where electron-electron interactions are neglected, or are taken into account at the mean-field level and absorbed into the electrostatic potential energy $\mathrm{V}_{\Gamma}(\bm{x})$ describing interactions between the electrons and the ions. Time-reversal symmetry is spontaneously broken by the inclusion of a static magnetic field
\begin{align}
    \bm{b}_{\mathrm{static}}(\bm{x}) = \bm{\nabla} \times \bm{a}_{\mathrm{static}}(\bm{x})
    \label{staticB}
\end{align}
having the periodicity of the crystal's lattice, meaning that $\bm{b}_{\mathrm{static}}(\bm{x} + \bm{R}_{\mathrm{t}}) = \bm{b}_{\mathrm{static}}(\bm{x})$ for all $\bm{R}_{\mathrm{t}} \in \Gamma$. Such a \textit{microscopic} magnetic field may arise, for example, due to local moments associated with one or more magnetic ions in each unit cell that persist in the system's ground state at zero temperature. Including only the leading order relativistic corrections coming from the Dirac Hamiltonian, we take the \textit{Hamiltonian density} to be
\begin{align}
    \mathcal{H}_0(\bm{x}) = \frac{1}{2m}\big(\bm{\mathfrak{p}}(\bm{x})\big)^2 + \mathrm{V}_{\Gamma}(\bm{x}) - \frac{e\hbar}{2mc} \bm{\sigma}\cdot\bm{b}_{\mathrm{static}}(\bm{x})\nonumber \\
    + \frac{\hbar}{4m^2 c^2} \bm{\sigma}\cdot\bm{\nabla}\mathrm{V}_{\Gamma}(\bm{x}) \times \bm{\mathfrak{p}}(\bm{x}),
    \label{H0density}
\end{align}
where under the standard minimal-coupling prescription, the (canonical) momentum operator is given by
\begin{align}
    \bm{\mathfrak{p}}(\bm{x}) \equiv \frac{\hbar}{i} \bm{\nabla} - \frac{e}{c} \bm{a}_{\mathrm{static}}(\bm{x}).
    \label{frakturp}
\end{align}
The first two terms of Eq. (\ref{H0density}) are to be thought of as being multiplied by the $2\times 2$ identity matrix, the second term is the Zeeman term, and the third term is the spin-orbit coupling term; we neglect additional relativistic corrections including the Darwin and mass-velocity terms.  Here we have introduced the $3$-tuple $\bm{\sigma} = (\sigma_1, \sigma_2, \sigma_3)$ of Pauli matrices forming a representation of the Lie algebra $\mathfrak{su}(2)$ associated with the nonrelativistic spin group $\mathrm{SU}(2)$, and, as with the magnetic field (\ref{staticB}) it generates, the static vector potential is cell-periodic.

Since the electrostatic potential energy, the static vector potential, and the magnetic field it generates are cell-periodic, so too is the Hamiltonian density (\ref{H0density}). Then Bloch's theorem \cite{Bloch} can be applied, and the solutions of the spectral equation
\begin{align}
    \mathcal{H}_0(\bm{x}) \psi_{n\bm{k}}(\bm{x}) = E_{n\bm{k}} \psi_{n\bm{k}}(\bm{x})
    \label{spectralpsi}
\end{align}
are the \textit{Bloch energy eigenfunctions}
\begin{align}
    \psi_{n\bm{k}}(\bm{x}) = \frac{1}{(2\pi)^{d/2}} e^{i\bm{k}\cdot\bm{x}} u_{n\bm{k}}(\bm{x}),
    \label{Blochtheorem}
\end{align}
where the cell-periodic parts $u_{n\bm{k}}(\bm{x})$ of the Bloch energy eigenfunctions, which we will refer to as \textit{Bloch functions}, satisfy the $\bm{k}$-dependent spectral equation
\begin{align}
    \mathcal{H}_{\bm{k}}(\bm{x}) u_{n\bm{k}}(\bm{x}) = E_{n\bm{k}} u_{n\bm{k}}(\bm{x}),
    \label{spectralu}
\end{align}
involving the modified Hamiltonian density
\begin{align}
    \mathcal{H}_0(\bm{x}) = \frac{1}{2m}\Big(\bm{\mathfrak{p}}(\bm{x}) + \hbar\bm{k}\Big)^2 + \mathrm{V}_{\Gamma}(\bm{x}) - \frac{e\hbar}{2mc} \bm{\sigma}\cdot\bm{b}_{\mathrm{static}}(\bm{x})\nonumber \\
    + \frac{\hbar}{4m^2 c^2} \bm{\sigma}\cdot\bm{\nabla}\mathrm{V}_{\Gamma}(\bm{x}) \times \Big(\bm{\mathfrak{p}}(\bm{x}) + \hbar\bm{k}\Big).
    \label{H0densityk}
\end{align}
The Bloch energy eigenfunctions and their corresponding Bloch functions are labelled by a band index $n \in \mathbb{N}$, while the Bloch wavevector $\bm{k}$ is a point in the Brillouin zone, which is a $d$-dimensional smooth manifold $\mathrm{BZ}^d$ that is diffeomorphic to the $d$-torus $\mathbb{T}^d$ \cite{Carpentier}, and the $2$-spinor Bloch functions can be written as \cite{Girvin}
\begin{align}
    u_{n\bm{k}}(\bm{x}) = \left(\mqty{u_{n\bm{k}}^{\uparrow}(\bm{x}) \\[1pt] u_{n\bm{k}}^{\downarrow}(\bm{x})}\right).
    \label{Blochfunctionspin}
 \end{align}
The Bloch energy eigenfunctions form a complete and orthonormal basis for the single-particle Hilbert space, and we can therefore expand the electron field operator in terms of these basis wavefunctions
\begin{align}
    \hat{\psi}_0(\bm{x},t) = \sum_{n=1}^{\infty} \int_{\mathrm{BZ}^d} d\bm{k}\, \psi_{n\bm{k}}(\bm{x}) \hat{a}_{n\bm{k}}(t),
    \label{electronfieldoperator}
\end{align}
where from Eq. (\ref{anticommutationrelations}) the electronic creation and annihilation operators satisfy the equal-time canonical anticommutation relations
\begin{align}
    \big[\hat{a}_{n\bm{k}}(t), \hat{a}_{m\bm{k}'}^{\dagger}(t)\big]_+ &=  \delta_{nm}\delta(\bm{k} - \bm{k}'),\nonumber \\
    \big[\hat{a}_{n\bm{k}}(t), \hat{a}_{m\bm{k}'}(t)\big]_+ &= \big[\hat{a}_{n\bm{k}}^{\dagger}(t), \hat{a}_{m\bm{k}'}^{\dagger}(t)\big]_+ = 0,
\end{align}
and the time evolution of these operators obtained from the Heisenberg equation (\ref{TDSE}) is given by
\begin{align}
    \hat{a}_{n\bm{k}}(t) = e^{-i E_{n\bm{k}}(t-t_0)/\hbar} \hat{a}_{n\bm{k}}(t_0).
\end{align}
The exact form of the Bloch energy eigenfunctions and the Bloch functions in Eqs. (\ref{spectralpsi}) and (\ref{spectralu}) depends on the crystal's lattice, and also on the electrostatic potential energy and the static vector potential in Eq. (\ref{H0density}).

To study the topology of band structure in crystals, it will be useful to introduce a set of abstract states $\ket{n\bm{k}}$ with $n \in \mathbb{N}$ for which the Bloch functions in Eq. (\ref{spectralu}) are $u_{n\bm{k}}(\bm{x}) = \bra{\bm{x}}\ket{n\bm{k}}$. These \textit{Bloch states} are collected into an ordered sequence $(\ket{n\bm{k}})_{n=1}^{\infty}$ that forms an orthonormal (Schauder) basis for a separable Hilbert space \cite{Kattan2}
\begin{align}
    \mathcal{B}_{\bm{k}} = \mathrm{span}_{\mathbb{C}}\big(\{\ket{n\bm{k}}\}_{n=1}^{\infty}\big)
    \label{BlochHilbertspace}
\end{align}
at each point $\bm{k} \in \mathrm{BZ}^d$, which is equipped with the Hermitian inner product
\begin{align}
    (f_{\bm{k}}|g_{\bm{k}}) = \frac{1}{\Omega_{\mathrm{uc}}} \int_{\Omega} d\bm{x}\, f_{\bm{k}}^{\dagger}(\bm{x}) g_{\bm{k}}(\bm{x})
    \label{innerproduct}
\end{align}
between any pair of elements $\ket{f_{\bm{k}}},\ket{g_{\bm{k}}} \in \mathcal{B}_{\bm{k}}$ with cell-periodic wavefunctions $f_{\bm{k}}(\bm{x})$ and $g_{\bm{k}}(\bm{x})$, where $\Omega$ is the unit cell and $\Omega_{\mathrm{uc}}$ is its volume. Importantly, if we consider mappings $\bm{k} \mapsto (\ket{n\bm{k}})_{n=1}^{\infty}$ assigning to each point $\bm{k}$ in some open set $U\subseteq \mathrm{BZ}^d$ this basis $(\ket{n\bm{k}})_{n=1}^{\infty}$ of Bloch states, then due to the presence of band crossings such mappings are not generally smooth (or even continuous). Indeed, if denote by $\mathscr{D}_{\mathcal{B}} \subseteq \mathrm{BZ}^d$ the locus of degeneracies in the crystal's band structure -- that is, the set of points $\bm{k}_0 \in \mathrm{BZ}^d$ for which $E_{n\bm{k}_0} = E_{m\bm{k}_0}$ for some pair of bands indexed by $n,m \in \mathbb{N}$ -- then these mappings $\bm{k} \mapsto (\ket{n\bm{k}})_{n=1}^{\infty}$, which we refer to as \textit{Bloch frames}, are only smooth over open sets $U \subseteq \mathrm{BZ}^d$ \textit{not} containing points in $\mathscr{D}_{\mathcal{B}}$ \cite{Kattan1}. However, we have shown \cite{CImanuscript1} that one can always form a collection of globally defined ``quasi-Bloch states" through $\bm{k}$-dependent unitary transformations of the Bloch states, from which a set of exponentially localized Wannier functions for a Chern insulator can be defined, albeit ones that include at least some of the unoccupied conduction bands as well as the occupied valence bands in their definition. The details of these constructions can be found in Appendix \ref{Appendix:QuasiBlochframes}. 

This is different from the more familiar situation of a \textit{topologically trivial insulator}, where exponentially localized Wannier functions (\ref{ELWF}) can be constructed from the occupied bands alone. Unlike in these topologically trivial insulators, for Chern insulators and other classes of topological insulators there are ``topological obstructions" to the definition of quantities like exponentially localized Wannier functions from the occupied bands alone, and these obstructions are encoded in certain \textit{topological invariants} that codify the way in which these occupied bands ``twist" across the Brillouin zone. The type of topological invariants that exist for a given system depends on its symmetry properties; for example, in the ``Class AII" of time-reversal symmetric $\mathbb{Z}_2$ topological insulators, the relevant topological invariants are $\mathbb{Z}_2$-valued indices. But for Chern insulators wherein time-reversal symmetry is broken, the relevant topological invariants are ``Chern number(s)" that are associated with the occupied valence bands. Collectively they will be referred to as \textit{Chern invariants}.

%%%%%%%%%%%%%%%%%%%%%%%%%%%%%%%%%%%%%%%%%%

\section{Chern invariants}\label{Sec:ChernInvariants}

Chern invariants characterize the topology of certain kinds of fibre bundles over the Brillouin zone. To define these invariants, we assemble the separable Hilbert spaces (\ref{BlochHilbertspace}) for all points $\bm{k} \in \mathrm{BZ}^d$ into a smooth Hilbert bundle $\mathcal{B}\overset{\pi_{\mathcal{B}}}{\longrightarrow} \mathrm{BZ}^d$, called the \textit{Bloch bundle} \cite{Panati2017}, such that the fibre above each point $\bm{k} \in \mathrm{BZ}^d$ is the Hilbert space (\ref{BlochHilbertspace}), all of which are isomorphic to a fixed Hilbert space $\mathcal{B}_0$ that is called the \textit{typical fibre} of $\mathcal{B}$. The Hermitian inner product (\ref{innerproduct}) on these Hilbert spaces smoothly extends to a Hermitian metric $h$ on the total space $\mathcal{B}$ of the Bloch bundle, assigning to each pair of smooth sections $\Psi,\Phi$ -- that is, to each pair of smooth maps $\Psi, \Phi : \mathrm{BZ}^d \to \mathcal{B}$ satisfying $\pi_{\mathcal{B}}(\Psi(\bm{k})) = \bm{k}$ and $\pi_{\mathcal{B}}(\Phi(\bm{k})) = \bm{k}$ for all points $\bm{k} \in \mathrm{BZ}^d$ of some open subset thereof -- a smooth function $h(\Psi,\Phi)$ on the $\mathrm{BZ}^d$ that is defined by \cite{Kattan2}
\begin{align}
    h(\Psi,\Phi)(\bm{k}) \equiv (\psi_{\bm{k}}|\varphi_{\bm{k}}),
\end{align}
where the right-hand-side is the inner product (\ref{innerproduct}) of a pair of abstract states $\ket{\psi_{\bm{k}}}$ and $\ket{\varphi_{\bm{k}}}$ in the fibre $\mathcal{B}_{\bm{k}}$ corresponding to the sections $\Psi(\bm{k})$ and $\Phi(\bm{k})$, respectively. The Bloch bundle includes Bloch states associated with \textit{all} of occupied and unoccupied bands, and is therefore a smooth Hilbert bundle of infinite rank, that is, its fibres are (countably) infinite-dimensional Hilbert spaces. There is an important theorem \cite{Freed} stating that every such infinite-rank smooth Hilbert bundle for which the fibres are separable Hilbert spaces is \textit{trivializable}, meaning that the Bloch bundle is isomorphic to the trivial bundle $\mathrm{BZ}^d \times \mathcal{B}_0$ involving the typical fibre $\mathcal{B}_0$ thereof; a particular such choice of isomorphism is called a \textit{trivialization} for $\mathcal{B}$ \cite{Panati2017}. That the Bloch bundle is trivializable is related to the fact that one can always find a collection globally defined and smooth quasi-Bloch functions from which exponentially localized Wannier functions for Chern insulators can be defined (see Appendix \ref{Appendix:QuasiBlochframes}).

Here we focus on \textit{band insulators} for which there is a well-defined bandgap separating the $N \in \mathbb{N}$ occupied valence bands from the remaining unoccupied conduction bands across the Brillouin zone. This means that we can define a smooth map $\bm{k} \mapsto P_{\mathcal{V}}(\bm{k})$ assigning to each point $\bm{k} \in \mathrm{BZ}^d$ an orthogonal projector $P_{\mathcal{V}}(\bm{k})$ that maps states in each fibre $\mathcal{B}_{\bm{k}}$ of the Bloch bundle onto the $N$-dimensional subspace $\mathcal{V}_{\bm{k}}$ thereof that is spanned by the Bloch states $\ket{n\bm{k}}$ associated with only the occupied valence bands $1 \leq n \leq N$. This is a finite-dimensional subspace of $\mathcal{B}_{\bm{k}}$ and therefore admits a well-defined orthogonal complement $\mathcal{C}_{\bm{k}}$ for each $\bm{k} \in \mathrm{BZ}^d$. Since the smooth map $\bm{k} \mapsto P_{\mathcal{V}}(\bm{k})$ is of fixed rank across the Brillouin zone, it induces a Whitney-sum decomposition of the Bloch bundle \cite{Kattan2}
\begin{equation}
    \mathcal{B} = \mathcal{V} \oplus \mathcal{C} \overset{\pi_{\mathcal{B}}}{\longrightarrow} \mathrm{BZ}^d,
    \label{Whitneysum}
\end{equation}
where $\mathcal{V} \overset{\pi_{\mathcal{V}}}{\longrightarrow} \mathrm{BZ}^d$ is called the \textit{valence bundle} and encodes the spectral properties of the occupied valence bands, while $\mathcal{C} \overset{\pi_{\mathcal{C}}}{\longrightarrow} \mathrm{BZ}^d$ is called the \textit{conduction bundle} and encodes the spectral properties of the unoccupied conduction bands \cite{Fruchart1}. Both of these are \textit{projected subbundles} of the Bloch bundle with fibres being defined through the projectors $P_{\mathcal{V}}(\bm{k})$ and $Q_{\mathcal{V}}(\bm{k}) \equiv \mathbb{I} - P_{\mathcal{V}}(\bm{k})$, and are also smooth Hilbert bundles over the $\mathrm{BZ}^d$.

While the Bloch bundle and its valence and conduction bundles contain the spectral properties of a crystal's band structure, its topological properties are described by their associated frame bundles: Denote by $F\mathcal{B}_{\bm{k}}$ the set of orthonormal Schauder bases for the fibre $\mathcal{B}_{\bm{k}}$ of the Bloch bundle $\mathcal{B}$ at each point $\bm{k}\in\mathrm{BZ}^d$. Examples include the basis of Bloch states $(\ket{n\bm{k}})_{n = 1}^{\infty}$ and any of bases formed from $\bm{k}$-dependent unitary transformations thereof, such as the quasi-Bloch states discussed in Appendix \ref{Appendix:QuasiBlochframes}. The sets $F\mathcal{B}_{\bm{k}}$ can be equipped with the structure of a smooth manifold, and are together assembled into a smooth fibre bundle over the $\mathrm{BZ}^d$ for which the total space is the disjoint union
\begin{equation}
    F\mathcal{B} \equiv \coprod_{\bm{k} \in \mathrm{BZ}^d} F\mathcal{B}_{\bm{k}},
    \label{framebundletotalspace}
\end{equation}
called the \textit{frame bundle} $F\mathcal{B} \overset{\pi}{\longrightarrow} \mathrm{BZ}^d$ associated to the Bloch bundle. Through standard constructions in differential geometry \cite{HamiltonBook,Kriegl}, this frame bundle can be equipped with the structure of a principal fibre bundle for which the structure group is the Lie group $\mathrm{U}(\mathcal{B}_0)$ of unitary operators on the typical fibre $\mathcal{B}_0$ of $\mathcal{B}$. Sections of the frame bundle $F\mathcal{B}$ are smooth frames over open sets in the Brillouin zone, examples of which include the locally-defined Bloch frames $\bm{k}\mapsto (\ket{n\bm{k}})_{n = 1}^{\infty}$. The frame bundle $F\mathcal{B}$ comes equipped with a right action of $\mathrm{U}(\mathcal{B}_0)$ that implements the unitary transformation (\ref{unitarytransformation}) between elements of these frames, as well as between different Wannier frames. We can similarly define the frame bundles $F\mathcal{V}$ and $F\mathcal{C}$ associated to the valence and conduction bundles, respectively, both of which are principal fibre bundles with the unitary structure groups $\mathrm{U}(\mathcal{V}_0)$ and $\mathrm{U}(\mathcal{C}_0)$, where $\mathcal{V}_0$ and $\mathcal{C}_0$ are the typical fibres of the valence and conduction bundles.

The Chern invariants describe how these fibre bundles ``twist" across the Brillouin zone, and we define these topological invariants within the context of Chern-Weil theory \cite{NakaharaBook}, that is, in terms of connections on these fibre bundles. Introduce a connection $1$-form $\omega$ on the frame bundle $F\mathcal{B}$, which is a $1$-form on the total space (\ref{framebundletotalspace}) that is valued in the Lie algebra $\mathfrak{u}(\mathcal{B}_0)$ of skew--self adjoint linear operators on $\mathcal{B}_0$. Since the Bloch bundle is trivializable as a smooth Hilbert bundle, there is a natural choice for this connection $1$-form that is defined in terms of the exterior derivative on the manifold $F\mathcal{B}$ \cite{Bohm1993}. To perform calculations using this connection, we introduce \textit{gauge potentials} through the pullback of $\omega$ by appropriate smooth sections of the frame bundle. The sections of interest are the local Bloch frames $\bm{k} \mapsto (\ket{n\bm{k}})_{n=1}^{\infty}$, where the gauge potential defined in terms of such a local Bloch frame is a locally defined Lie algebra--valued $1$-form
\begin{align}
    \xi(\bm{k}) &= \sum_{mn} \xi_{mn}(\bm{k}) \ketbra{m\bm{k}}{n\bm{k}},
    \label{BerryconnectionB}
\end{align}
where the sum includes all of the bands and the corresponding matrix-valued $1$-form is given by \cite{Kattan2}
\begin{align}
    \xi_{mn}(\bm{k}) = i(m\bm{k}|\mathrm{d}n\bm{k}),
    \label{BerryconnectionBcomponents}
\end{align}
involving the exterior derivative $\mathrm{d} = \mathrm{d}k^a \partial / \partial k^a$ and the inner product (\ref{innerproduct}). Because this quantity involves Bloch-frame matrix elements, it is generally only smooth on open sets not containing points in the locus of degeneracies $\mathscr{D}_{\mathcal{B}}$. Working in a Cartesian chart for the $\mathrm{BZ}^d$, the Cartesian components of the non-Abelian Berry connection in this Bloch frame are \cite{VanderbiltBook}
\begin{align}
    \xi_{mn}^i(\bm{k}) = \frac{i}{\Omega_{\mathrm{uc}}} \int_{\Omega} d\bm{x}\, u_{m\bm{k}}^{\dagger}(\bm{x}) \frac{\partial u_{n\bm{k}}(\bm{x})}{\partial k^i}.
\end{align}
The Berry curvature of the local Bloch-frame gauge potential (\ref{BerryconnectionB}) is the Lie algebra--valued $2$-form
\begin{align}
    F(\bm{k}) &= \mathrm{d}\xi(\bm{k}) - i \xi(\bm{k}) \wedge \xi(\bm{k}),
    \label{BerrycurvatureB}
\end{align}
which through an expansion analogous to Eq. (\ref{BerryconnectionB}) can be written in terms of the local matrix-valued $2$-form
\begin{align}
    F_{mn}(\bm{k}) = \mathrm{d}\xi_{mn}(\bm{k}) - i \sum_{\ell} \xi_{m\ell}(\bm{k}) \wedge \xi_{\ell n}(\bm{k}).
    \label{BerrycurvatureBcomponents}
\end{align}
Again, this quantity involves Bloch-frame matrix elements and is therefore not generally smooth at points in the locus of degeneracies. Now, since the local gauge potential (\ref{BerryconnectionB}) is obtained from a connection $1$-form $\omega$ on $F\mathcal{B}$ that is ``flat," one can show that the local curvature $2$-form (\ref{BerrycurvatureB}) vanishes \cite{Bohm1993}; that is, we have 
\begin{align}
    F(\bm{k}) = 0 \implies \mathrm{d}\xi(\bm{k}) = i \xi(\bm{k}) \wedge \xi(\bm{k}).
    \label{BerrycurvatureBzero}
\end{align}
Working in a Cartesian chart, the chart representation of this result leads to the local Bloch-frame identity \cite{CImanuscript1}
\begin{align}
    \varepsilon^{iab} \partial_a \xi_{mn}^b(\bm{k}) &= i \sum_{\ell} \varepsilon^{iab} \xi_{m\ell}^a(\bm{k}) \xi_{\ell n}^b(\bm{k}),
    \label{CurvaturefreeBlochframe}
\end{align}
valid away from points in the locus of degeneracies $\mathscr{D}_{\mathcal{B}}$. 

Recall that the valence bundle $\mathcal{V}$ is a rank-$N$ subbundle of the Bloch bundle $\mathcal{B}$ that encodes the spectral properties of the occupied ``valence" bands in an insulator's band structure. Because the valence bundle is a \textit{projected} subbundle, the connection $1$-form on its frame bundle $F\mathcal{V}$ is the projected connection $\omega_{\mathcal{V}} = P_{\mathcal{V}}\omega$ of the connection $\omega$ on $F\mathcal{B}$. Then the local gauge potential for $\omega_{\mathcal{V}}$ defined through a local Bloch frame $\bm{k} \mapsto (\ket{n\bm{k}})_{n=1}^{N}$ involves only the Bloch states for the occupied valence bands,
\begin{align}
    \xi_{\mathcal{V}}(\bm{k}) = \sum_{mn}^{N} \xi_{\mathcal{V},mn}(\bm{k}) \ketbra{m\bm{k}}{n\bm{k}},
    \label{BerryconnectionV}
\end{align}
where the matrix-valued $1$-form is given by
\begin{align}
    \xi_{\mathcal{V},mn}(\bm{k}) = i (m\bm{k}|P_{\mathcal{V}}(\bm{k})\mathrm{d}n\bm{k}).
    \label{BerryconnectionVcomponents}
\end{align}
When expanded in terms of the Bloch states, the projector in this expression restricts the band indices to only the occupied ones $1\leq n,m \leq N$. The curvature of this gauge potential is the Lie algebra--valued $2$-form
\begin{align}
    F_{\mathcal{V}}(\bm{k}) = \mathrm{d}\xi_{\mathcal{V}}(\bm{k}) - i \xi_{\mathcal{V}}(\bm{k})\wedge \xi_{\mathcal{V}}(\bm{k}),
    \label{BerrycurvatureV}
\end{align}
and, introducing an analogous Bloch-frame expansion to Eq. (\ref{BerryconnectionV}), the corresponding matrix-valued $2$-form is
\begin{align}
    F_{\mathcal{V},mn}(\bm{k}) = \mathrm{d}\xi_{\mathcal{V},mn}(\bm{k}) - i \sum_{\ell = 1}^N \xi_{\mathcal{V},m\ell}(\bm{k})\wedge \xi_{\mathcal{V},\ell n}(\bm{k}),
    \label{BerrycurvatureVcomponents}
\end{align}
valid away from points of degeneracy. While the connection on the Bloch bundle has vanishing curvature, this is no longer true for the projected connection on the valence bundle. Indeed, the spectral projector $\bm{k} \mapsto P_{\mathcal{V}}(\bm{k})$ can induce nontrivial curvature on $F\mathcal{V}$ and is ultimately responsible for the topological structure of the valence bundle, as can be seen by writing the curvature $2$-form (\ref{BerrycurvatureV}) in terms of these projectors \cite{Brouder},
\begin{align}
    F_{\mathcal{V}}(\bm{k}) = i P_{\mathcal{V}}(\bm{k}) \mathrm{d}P_{\mathcal{V}}(\bm{k})\wedge\mathrm{d}P_{\mathcal{V}}(\bm{k}) P_{\mathcal{V}}(\bm{k}),
    \label{projectorBerrycurvatureV}
\end{align}
which is globally defined across the $\mathrm{BZ}^d$. While we could define the Chern invariants below in terms of this globally smooth expression for the Berry curvature, the trouble with this formula is that when performing any numerical calculations of the Berry curvature or the Chern invariants, one inevitably needs to introduce a basis expansion of the projectors, after which one again ends up with the locally defined expression (\ref{BerrycurvatureVcomponents}).

\subsection{The Chern number in two dimensions}\label{Sec:ChernNumbers}

Now consider a Chern insulator in $d = 2$ dimensions. Here the relevant topological invariants are the first Chern numbers of the Bloch bundle and its valence bundle over the two-dimensional Brillouin zone $\mathrm{BZ}^2$ \cite{Monaco2017}. The first Chern number $C_{\mathcal{B}}$ for the Bloch bundle $\mathcal{B}$ can be obtained by integrating a certain characteristic class over the $\mathrm{BZ}^2$, called the \textit{first Chern class} \cite{NakaharaBook}
\begin{align}
    \mathrm{c}_1\big(F(\bm{k})\big) \equiv \frac{1}{2\pi} \tr_{\mathcal{B}}\big(F(\bm{k})\big),
    \label{CherncharacterB}
\end{align}
where the trace involves a sum over all of the occupied \textit{and} unoccupied bands in the crystal's band structure, and the first Chern number for the Bloch bundle is 
\begin{align}
    C_{\mathcal{B}} = \int_{\mathrm{BZ}^2} \mathrm{c}_1\big(F(\bm{k})\big).
    \label{ChernnumberB}
\end{align}
But the first Chern class (\ref{CherncharacterB}) clearly vanishes, since the curvature $2$-form satisfies Eq. (\ref{BerrycurvatureBzero}), and it follows that this first Chern number $C_{\mathcal{B}} = 0$ \cite{CImanuscript1}. This result is independent of our choice of connection $1$-form on the frame bundle, a consequence of the Chern-Weil theorem \cite{NakaharaBook}, and it follows that the Bloch bundle $\mathcal{B}$ being trivializable is equivalent to the first Chern number $C_{\mathcal{B}}$ vanishing. 

For a Chern insulator it isn't the Bloch bundle $\mathcal{B}$ that is ``topologically nontrivial," but rather its valence bundle $\mathcal{V}$ describing only the occupied valence bands. The first Chern class for the valence bundle is given in terms of the curvature $2$-form (\ref{BerrycurvatureV}) by 
\begin{align}
    \mathrm{c}_1\big(F_{\mathcal{V}}(\bm{k})\big) = \frac{1}{2\pi} \tr_{\mathcal{V}}\big(F_{\mathcal{V}}(\bm{k})\big),
    \label{CherncharacterV}
\end{align}
where the trace is over only the occupied valence bands, and the first Chern number of the valence bundle is
\begin{align}
    C_{\mathcal{V}} = \int_{\mathrm{BZ}^2} \mathrm{c}_1\big(F_{\mathcal{V}}(\bm{k})\big),
    \label{ChernnumberV}
\end{align}
which is valued in the integers $\mathbb{Z}$ and \textit{does not} vanish for a Chern insulator \cite{Monaco2017}. Working in a Cartesian chart with local coordinates $\bm{k} = (k_x, k_y)$, one can show that the chart representation of this Chern number is given by the well-known expression \cite{VanderbiltBook}
\begin{align}
    C_{\mathcal{V}} = \frac{1}{2\pi} \sum_n f_n \int_{\mathrm{BZ}^2} d\bm{k}\, \varepsilon^{ab} \partial_a \xi_{nn}^b(\bm{k}),
    \label{ChernnumberV2}
\end{align}
involving the Cartesian components of the matrix-valued $1$-form (\ref{BerryconnectionVcomponents}) \footnote{To properly define this expression, a small region containing each point in the locus of degeneracies must be removed, and then the contributions of these regions to the integral over the $\mathrm{BZ}^d$ must be evaluated using the generalized Stokes theorem \cite{Kattan2}\label{foot1}}. Here we have defined the Levi-Civita symbol with components $\varepsilon^{xy} = - \varepsilon^{yx} = 1$ and $\varepsilon^{xx} = \varepsilon^{yy} = 0$, and the filling factor $f_n = 1$ if $n$ labels an occupied valence band, and $f_n = 0$ otherwise.

Any calculation of the Chern number using the chart representation (\ref{ChernnumberV2}) will involve evaluating diagonal components of the Berry connection, which can be problematic due to the presence of band crossings, especially when an integral over the Brillouin zone is involved. An advantage of the formalism introduced in Sec. \ref{Sec:HamiltonianTheory} is that we can derive a \textit{global expression} for this Chern number, that is, a chart representation for which the integrand is still globally defined and smooth. Noting that the velocity operator obtained from the Hamiltonian (\ref{H0density}) is
\begin{align}
    \bm{v}(\bm{x}) = \frac{1}{m} \bm{\mathfrak{p}}(\bm{x}) + \frac{\hbar}{4m^2 c^2} \bm{\sigma}\times \bm{\nabla}\mathrm{V}_{\Gamma}(\bm{x}),
    \label{velocityoperator}
\end{align}
it is straightforward to show using the identity \cite{CImanuscript1}
\begin{align}
    v_{nm}^i(\bm{k}) = \frac{\delta_{nm}}{\hbar} \frac{\partial E_{m\bm{k}}}{\partial k^i} + \frac{i}{\hbar} (E_{n\bm{k}} - E_{m\bm{k}}) \xi_{nm}^i(\bm{k}),
    \label{velocityBerryconnection}
\end{align}
that this global expression is given by
\begin{align}
    C_{\mathcal{V}} = \frac{h^2}{4\pi i} \sum_{mn} f_{nm} \int_{\mathrm{BZ}^2} \frac{d\bm{k}}{(2\pi)^2} \frac{\varepsilon^{ab} v_{nm}^a(\bm{k}) v_{mn}^b(\bm{k})}{(E_{n\bm{k}} - E_{m\bm{k}})(E_{m\bm{k}} - E_{n\bm{k}})},
    \label{microscopicChernnumberV}
\end{align}
involving the matrix elements
\begin{align}
    \bm{v}_{nm}(\bm{k}) = \frac{1}{\Omega_{\mathrm{uc}}} \int_{\Omega} d\bm{x}\, u_{n\bm{k}}^{\dagger}(\bm{x}) \bm{v}(\bm{x}) u_{m\bm{k}}(\bm{x}).
    \label{velocitymatrixelements}
\end{align}
The expression (\ref{microscopicChernnumberV}) is well-behaved when $m = n$ as the prefactor $f_{nm} = f_{n} - f_m$ involving the filling factors $f_{n},f_{m}$ vanish, so that the poles in the denominator do not contribute and any issues regarding degeneracies across the Brillouin zone are not present.

\subsection{The Chern vector in three dimensions}\label{Sec:ChernVector}

Consider a Chern insulator in $d = 3$ dimensions. The topological invariants for Chern insulators in three dimensions are more subtle to define than the first Chern number (\ref{ChernnumberV}) in two dimensions. There are now a triple of first Chern numbers $C_{\mathcal{V}}^{\alpha} \in \mathbb{Z}$ indexed by $1\leq \alpha \leq 3$ characterizing the topological structure of the valence bundle $\mathcal{V}$ over the three-dimensional Brillouin zone \cite{Troyer2016, Panati2017}, which are defined by integrating the first Chern class (\ref{CherncharacterV}) over the three $2$-cycles of the  $\mathrm{BZ}^3 \cong \mathbb{T}^3$. These Chern numbers are assembled into the \textit{Chern vector} \cite{Kattan2}
\begin{align}
    \bm{C}_{\mathcal{V}} = \sum_{\alpha = 1}^3 C_{\mathcal{V}}^{\alpha}\, \bm{g}_{\alpha},
    \label{Chernvector}
\end{align}
which is a vector-valued topological invariant valued in the crystal's reciprocal lattice. To define these Chern numbers, we decompose the $\mathrm{BZ}^3$ into its three $2$-cycles \cite{CImanuscript1}, which are embedded submanifolds that are diffeomorphic to the $2$-torus $\mathbb{T}^2$ and are defined through a triple of smooth embedding maps $\phi_{\alpha} : \mathbb{T}^2 \hookrightarrow \mathrm{BZ}^3$ for each $1 \leq \alpha \leq 3$; the $\alpha\mathrm{th}$ $2$-cycle is the image $\mathbb{T}_{\alpha}^2 \equiv \phi_{\alpha}(\mathbb{T}^2)$ under the corresponding embedding map. Then the first Chern number $C_{\mathcal{B}}^{\alpha}$ of the Bloch bundle $\mathcal{B}$ is the integral of the first Chern class (\ref{CherncharacterB}) over the $\alpha\mathrm{th}$ 2-cycle, 
\begin{align}
    C_{\mathcal{B}}^{\alpha} = \int_{\mathbb{T}_{\alpha}^2} \mathrm{c}_1\big(F(\bm{k})\big),
    \label{Chernnumbers3DB}
\end{align}
and, since the curvature $2$-form (\ref{BerrycurvatureB}) for the Bloch bundle vanishes, it follows that $C_{\mathcal{B}}^{\alpha} = 0$ for all $1 \leq \alpha \leq 3$. But for a Chern insulator at least one of the first Chern numbers $C_{\mathcal{V}}^{\alpha}$ of the valence bundle $\mathcal{V}$ will be nonzero, which are defined to be the integrals of the first Chern class (\ref{CherncharacterV}) of the valence bundle over the same $2$-cycles,
\begin{align}
    C_{\mathcal{V}}^{\alpha} = \int_{\mathbb{T}_{\alpha}^2} \mathrm{c}_1\big(F_{\mathcal{V}}(\bm{k})\big),
    \label{Chernnumbers3DV}
\end{align}
all three of which are valued in the integers. Working in a Cartesian chart with local coordinates $\bm{k} = (k_x, k_y, k_z)$, one can show that the chart representation of these Chern numbers is given by \cite{Kattan2}
\begin{align}
    C_{\mathcal{V}}^{\alpha} = \frac{1}{2\pi} H^{\alpha}_{\;\; a}  \sum_n f_n \int_{\mathrm{BZ}^3} d\bm{k}\, \varepsilon^{abc} \partial_{b} \xi_{nn}^c(\bm{k}),
    \label{CV3Dchart}
\end{align}
involving the quantities
\begin{align}
    H_{\;\; a}^{\alpha} = \frac{1}{2\pi} \delta^{\alpha\beta} \bm{a}_{\beta} \cdot \hat{\bm{e}}_a,
    \label{Hmatrix}
\end{align}
where $\bm{a}_{\beta}$ is a primitive lattice vector and $\hat{\bm{e}}_a$ is the $a\mathrm{th}$ Cartesian unit vector \footnote{See [\hyperref[foot1]{52}].}. It should be emphasized that it may be the case that two or even all three of these first Chern numbers are nonzero, but only one of them needs to be nonzero for a system to be a Chern insulator.

As in the case of the Chern number (\ref{ChernnumberV2}), the integrand is only defined locally away from points of degeneracy. But we can again find a \textit{global expression} for each of these Chern numbers (\ref{Chernnumbers3DV}). Using the identity (\ref{velocityBerryconnection}) for the velocity matrix elements, we find
\begin{widetext}
\begin{align}
    C_{\mathcal{V}}^{\alpha} = \frac{h^2}{4\pi i} H_{\;\; a}^{\alpha} \sum_{mn} f_{nm} \int_{\mathrm{BZ}^3} \frac{d\bm{k}}{(2\pi)^3}\, \frac{\varepsilon^{abc} v_{nm}^b(\bm{k})v_{mn}^c(\bm{k})}{(E_{n\bm{k}} - E_{m\bm{k}})(E_{m\bm{k}} - E_{n\bm{k}})}.
    \label{microscopicChernnumberV3D}
\end{align}
\end{widetext}
While these Chern numbers clearly depend on our choice of primitive lattice vectors, the Chern vector does not depend on this choice. Indeed, after combining Eqs. (\ref{Chernvector}) and (\ref{microscopicChernnumberV3D}) it is straightforward to show that the Chern vector can be written
\begin{align}
    \bm{C}_{\mathcal{V}} = \frac{h^2}{4\pi i} \sum_{mn} f_{nm} \int_{\mathrm{BZ}^3}\frac{d\bm{k}}{(2\pi)^3}\frac{\hat{\bm{e}}_a \varepsilon^{abc} v_{nm}^b(\bm{k}) v_{mn}^c(\bm{k})}{(E_{n\bm{k}} - E_{m\bm{k}})(E_{m\bm{k}} - E_{n\bm{k}})},
    \label{microscopicChernvector}
\end{align}
which does not depend on our choice of primitive lattice vectors for the crystal's lattice (\ref{lattice}). While the quantity $\hat{\bm{e}}_{i}$ is a Cartesian unit vector, the velocity matrix elements $v_{nm}^i(\bm{k})$ in the numerator are the \textit{Cartesian components} of the velocity operator (\ref{velocityoperator}), and so this is indeed a vector-valued quantity that is invariant under proper rotations of the Cartesian unit vectors. In fact, this expression can be shown to be invariant under general coordinate transformations (chart transition maps) in the three-dimensional Brillouin zone $\mathrm{BZ}^3$. Heuristically, this follows from the fact that the numerator in the integrand of the Chern vector (\ref{microscopicChernvector}) effectively involves the ``cross product" $\bm{\mathfrak{v}}_{nm}(\bm{k}) \times \bm{\mathfrak{v}}_{mn}(\bm{k})$ of the vector fields associated with the velocity matrix elements.

%%%%%%%%%%%%%%%%%%%%%%%%%%%%%%%%%%%%%%%%%%

\section{Linear response}\label{Sec:LinearResponse}

The Chern invariants discussed in Sec. \ref{Sec:ChernInvariants} are properties of the unperturbed band structure of a Chern insulator \cite{CImanuscript1}. We now consider the scenario where there is an electromagnetic field is driving the system, in which case the field theory in Sec. \ref{Sec:HamiltonianTheory} needs to be modified \cite{Mahon2019}. Instead of the field operator $\hat{\psi}_0(\bm{x},t)$ for the unperturbed medium, in the presence of an electromagnetic field the electronic degrees of freedom are now described by an electron field operator $\hat{\psi}(\bm{x},t)$ obeying the same anticommutation relations (\ref{anticommutationrelations}), where the dynamical evolution of this field operator is governed by a Heisenberg equation with the second-quantized Hamiltonian 
\begin{align}
    \hat{H}_{\mathrm{mc}}(t) = \int d\bm{x}\, \hat{\psi}^{\dagger}(\bm{x},t)\mathcal{H}_{\mathrm{mc}}(\bm{x},t)\hat{\psi}(\bm{x},t),
\end{align}
involving the minimal-coupling Hamiltonian density
\begin{align}
    \mathcal{H}_{\mathrm{mc}}(\bm{x},t) = \frac{1}{2m}\left(\bm{\mathfrak{p}}(\bm{x}) - \frac{e}{c}\bm{A}(\bm{x},t)\right)^2 + \mathrm{V}_{\Gamma}(\bm{x}) + e\phi(\bm{x},t)\nonumber \\
    + \frac{\hbar}{4m^2 c^2} \bm{\sigma}\cdot\bm{\nabla}\mathrm{V}_{\Gamma}(\bm{x}) \times \left(\bm{\mathfrak{p}}(\bm{x}) - \frac{e}{c}\bm{A}(\bm{x},t)\right)\nonumber \\
    - \frac{e\hbar}{2mc} \bm{\sigma}\cdot\bm{b}_{\mathrm{static}}(\bm{x}) - \frac{e\hbar}{2mc} \bm{\sigma}\cdot\bm{B}(\bm{x},t)
    \label{Hmcdensity}
\end{align}
Here we neglect local field corrections, which are associated with the relation between the microscopic electromagnetic field and its macroscopic counterpart, taking the associated electric and magnetic fields to be the macroscopic Maxwell fields, which are described in the Hamiltonian density in terms of the scalar potential $\phi(\bm{x},t)$ and vector potential $\bm{A}(\bm{x},t)$ \cite{Jackson}.

\subsection{Conductivity tensor}\label{Sec:ConductivityTensor}

We consider here the long-wavelength response of a bulk Chern insulator in the linear regime. Generalizing our previous work \cite{Kattan2} on the long-wavelength response to include the spin degree of freedom \cite{Kattan2}, the macroscopic current density of a Chern insulator in $d = 2$ dimensions ($i,\ell \in \{x,y\}$) is 
\begin{align}
    K^i(\bm{x},\omega) = \sigma^{i\ell}(\omega) E_{\parallel}^{\ell}(\bm{x},\omega),
    \label{sheetcurrentdensity}
\end{align}
where $E_{\parallel}^i(\bm{x},\omega)$ are the components of the total electric field tangential to the sheet, and the two-dimensional conductivity tensor is given by
\begin{align}
    \sigma^{i\ell}(\omega) = \sigma_{\mathrm{K}}^{i\ell}(\omega) - \frac{e^2}{2\pi\hbar} \varepsilon^{i\ell} C_{\mathcal{V}}
    \label{totalconductivity2D}
\end{align}
involving the Chern number (\ref{microscopicChernnumberV}). For a Chern insulator in $d = 3$ dimensions ($i,\ell \in \{x,y,z\}$), the macroscopic current density is 
\begin{align}
    J^i(\bm{x},\omega) = \sigma^{i\ell}(\omega) E^{\ell}(\bm{x},\omega),
    \label{bulkcurrentdensity}
\end{align}
where $E^{i}(\bm{x},\omega)$ is the total electric field, and the three-dimensional conductivity tensor is given by
\begin{align}
    \sigma^{i\ell}(\omega) = \sigma_{\mathrm{K}}^{i\ell}(\omega) - \frac{e^2}{2\pi\hbar} \varepsilon^{i\ell j}\, \hat{\bm{e}}_j \cdot\bm{C}_{\mathcal{V}},
    \label{totalconductivity3D}
\end{align}
involving the Chern vector (\ref{microscopicChernvector}). Note that the conductivities have different units in two and three dimensions. But in both cases, the first term is a \textit{Kubo conductivity tensor} that takes the same form as the conductivity tensor that would be obtained from a Kubo analysis for a topologically trivial insulator \cite{Mahon2020a}, and is given by
\begin{widetext}
\begin{align}
    \sigma_{\mathrm{K}}^{i\ell}(\omega) = - i\omega e^2 \hbar^2 \sum_{mn} f_{nm} \int_{\mathrm{BZ}^d} \frac{d\bm{k}}{(2\pi)^d} \frac{1}{(E_{m\bm{k}} - E_{n\bm{k}})^2}\frac{v_{nm}^{i}(\bm{k}) v_{mn}^{\ell}(\bm{k})}{E_{m\bm{k}} - E_{n\bm{k}} - \hbar(\omega + i0^+)}.
    \label{Kuboconductivity}
\end{align}
\end{widetext}
while the second terms in Eqs. (\ref{totalconductivity2D}) and (\ref{totalconductivity3D}) are the Hall conductivity tensors in two and three dimensions. The Kubo conductivity tensor vanishes in the static limit $(\omega \to 0)$, and the Hall conductivity tensor that remains describes the quantum anomalous Hall effect \cite{QAHcolloquium}. This global expression for the Kubo conductivity tensor shows explicitly how the usual Kubo response of a topologically trivial insulator is modified when a time-reversal breaking vector potential is included in the Hamiltonian (\ref{H0density}), namely in the modification of the matrix elements of the velocity operator (\ref{velocityoperator}). Notably, we show in Appendix \ref{Appendix:KuboGreenwood} that for Chern insulators in both two and three dimensions the Kubo-Greenwood formula \cite{LeiMacdonald}, given by 
\begin{widetext}
\begin{align}
    \sigma_{\mathrm{KG}}^{i\ell}(\omega) = \frac{ie^2}{\hbar} \sum_{mn} f_{mn} \int_{\mathrm{BZ}^d}\frac{d\bm{k}}{(2\pi)^d} \frac{1}{E_{m\bm{k}} - E_{n\bm{k}}} \frac{(n\bm{k}|\partial_{i}H_{\bm{k}}|m\bm{k})(m\bm{k}|\partial_{\ell}H_{\bm{k}}|n\bm{k})}{E_{m\bm{k}}-E_{n\bm{k}} - \hbar(\omega + i 0^+)},
    \label{KuboGreenwood}
\end{align}
\end{widetext}
is in agreement with Eqs. (\ref{totalconductivity2D}) and (\ref{totalconductivity3D}). The integrand above features the matrix elements
\begin{align}
    (n\bm{k}|\partial_{i}H_{\bm{k}}|m\bm{k}) = \frac{1}{\Omega_{\mathrm{uc}}} \int_{\Omega} d\bm{x}\, u_{n\bm{k}}^{\dagger}(\bm{x}) \partial_i \mathcal{H}_{\bm{k}}(\bm{x}) u_{m\bm{k}}(\bm{x})
    \label{matrixpartialH}
\end{align}
of a generic $\bm{k}$-dependent Hamiltonian. If we were considering a topologically trivial insulator for which the static vector potential in the Hamiltonian density (\ref{H0densityk}) vanishes, then this Kubo-Greenwood formula reduces to the Kubo conductivity tensor (\ref{Kuboconductivity}). However, when the static vector potential is included in this Hamiltonian density, then a more careful analysis of the Kubo-Greenwood formula is necessary. Before the discovery of topological materials, it was often implicitly assumed that time-reversal symmetry holds in bulk materials, in which case no such Hall conductivity would arise \cite{Kubo,Greenwood}. But it is now understood that there exist materials, including Chern insulators, for which time-reversal symmetry is broken, and we show in Appendix \ref{Appendix:KuboGreenwood} that the Kubo-Greenwood formula (\ref{KuboGreenwood}) yields the same conductivity tensors (\ref{totalconductivity2D}) and (\ref{totalconductivity3D}) when the Hamiltonian (\ref{H0densityk}) of our field theory is used in Eq. (\ref{matrixpartialH}), so that $\sigma_{\mathrm{KG}}^{i\ell}(\omega) = \sigma^{i\ell}(\omega)$ in both two and three dimensions.

\subsection{Effective dielectric tensor}\label{Sec:DielectricTensor}

When studying the long-wavelength optics of bulk materials, it is common to introduce an \textit{effective dielectric tensor} relating the displacement field in the medium to the electric field \cite{Bakry}, which is given by
\begin{align}
    \varepsilon^{i\ell}(\omega) = \delta^{i\ell} + \frac{4\pi i}{\omega} \sigma^{i\ell}(\omega).
\end{align}
We can then use our conductivity tensor (\ref{totalconductivity3D}) to obtain the effective dielectric tensor for a bulk Chern insulators in $d = 3$ dimensions,
\begin{align}
    \varepsilon^{i\ell}(\omega) = \delta^{i\ell} + 4\pi \chi^{i\ell}(\omega) + \frac{2ie^2}{\hbar \omega} \varepsilon^{ij\ell}\, \hat{\bm{e}}_j \cdot \bm{C}_{\mathcal{V}},
    \label{effectivedielectrictensor}
\end{align}
where we have introduced a \textit{Kubo susceptibility tensor}
\begin{widetext}
\begin{align}
    \chi^{i\ell}(\omega) = e^2 \hbar^2 \sum_{mn} f_{nm} \int_{\mathrm{BZ}^3} \frac{d\bm{k}}{(2\pi)^3} \frac{1}{(E_{m\bm{k}} - E_{n\bm{k}})^2} \frac{v_{nm}^i(\bm{k}) v_{mn}^{\ell}(\bm{k})}{E_{m\bm{k}} - E_{n\bm{k}} - \hbar(\omega + i0^+)}.
    \label{Kubosusceptibility}
\end{align}
\end{widetext}
There are several interesting properties of the effective dielectric tensor (\ref{effectivedielectrictensor}) that are not present for time-reversal symmetric and topologically trivial insulators. For example, the anisotropy of both the Kubo susceptibility tensor and the Hall conductivity leads to a pair of distinct refractive indices that are associated with the left and right circular polarization components of light propagating in a Chern insulator. This means that a Chern insulator is optically active, displaying both circular birefringence and circular dichroism, which are encoded in the Faraday and Kerr angles for the transmitted and reflected fields across a finite-sized sample, respectively. In a forthcoming publication we will calculate these angles for light propagating across a thin film of the Chern insulator $\mathrm{Mn}\mathrm{Bi}_2\mathrm{Te}_4$ with seven septuple layers. 

To better understand which properties of this dielectric tensor for Chern insulators are not present in topologically trivial and time-reversal invariant media, it is useful to study the causal properties of this tensor, which are encoded the generalized Kramers-Kronig relations \cite{lucarini2005kramers, Bakry} 
\begin{widetext}
\begin{align}
    \mathscr{R}\big(\varepsilon^{i\ell}(\omega) - \delta^{i\ell}\big) &= \frac{1}{\pi}\, \fint_{-\infty}^{\infty} d\omega'\, \frac{\mathscr{A}\big(\varepsilon^{i\ell}(\omega') - \delta^{i\ell}\big)}{\omega' - \omega} + \frac{2i e^2}{\hbar\omega} \varepsilon^{ij\ell}\,\hat{\bm{e}}_j \cdot \bm{C}_{\mathcal{V}},\nonumber \\
    \mathscr{A}\big(\varepsilon^{i\ell}(\omega) - \delta^{i\ell}\big) &= - \frac{1}{\pi}\fint_{-\infty}^{\infty} d\omega'\, \frac{\mathscr{R}\big(\varepsilon^{i\ell}(\omega') - \delta^{i\ell}\big)}{\omega' - \omega},
\end{align}
\end{widetext}
involving Cauchy principal value integrals, indicated here and below by bars on the integrals, where we have used the fact that the Hall contribution to the dielectric tensor has no absorptive part. One can show that this implies that the Kubo susceptibility tensor satisfies the usual Kramers-Kronig relations \cite{Jackson}. The \textit{reactive} and \textit{absorptive} parts of the effective dielectric tensor are defined by
\begin{align}
    \mathscr{R}\big(\varepsilon^{i\ell}(\omega)\big) &\equiv \frac{1}{2} \Big(\varepsilon^{i\ell}(\omega) + \big(\varepsilon^{\ell i}(\omega)\big)^*\Big),\nonumber \\
    \mathscr{A}\big(\varepsilon^{i\ell}(\omega)\big) &\equiv \frac{1}{2i} \Big(\varepsilon^{i\ell}(\omega) - \big(\varepsilon^{\ell i}(\omega)\big)^*\Big),
    \label{reactiveabsorptive}
\end{align}
from which follows
\begin{align}
    \varepsilon^{i\ell}(\omega) = \mathscr{R}\big(\varepsilon^{i\ell}(\omega)\big) + i\mathscr{A}\big(\varepsilon^{i\ell}(\omega)\big).
\end{align}
The reactive and absorptive parts of the static part of the conductivity tensor are defined similarly. For time-reversal invariant systems, the Kubo susceptibility tensor (\ref{Kubosusceptibility}) can be shown to satisfy $\chi^{i\ell}(\omega) = \chi^{\ell i}(\omega)$, in which case the reactive and absorptive parts (\ref{reactiveabsorptive}) are just the real and imaginary parts of this tensor. This is no longer the case for any system in which time-reversal symmetry is broken, including topologically trivial materials with long-range magnetic order, along with Chern insulators. 

We can find explicit formulae for the reactive and absorptive parts (\ref{reactiveabsorptive}) using the standard identity \cite{Jackson}
\begin{align}
    \lim_{\varepsilon \to 0^+} \frac{1}{\omega' - \omega \pm i\varepsilon} = \mathrm{p.v.}\left(\frac{1}{\omega' - \omega}\right) \mp i \pi \delta(\omega' - \omega)
\end{align}
in the Kubo susceptibility tensor (\ref{Kubosusceptibility}), in which case the reactive and absorptive parts thereof are given by
\begin{widetext}
\begin{align}
    \mathscr{R}\big(\chi^{i\ell}(\omega)\big) &= e^2 \hbar^2 \sum_{mn} f_{nm} \fint_{\mathrm{BZ}^3} \frac{d\bm{k}}{(2\pi)^3} \frac{1}{(E_{m\bm{k}} - E_{n\bm{k}})^2} \frac{v_{nm}^i(\bm{k}) v_{mn}^{\ell}(\bm{k})}{E_{m\bm{k}} - E_{n\bm{k}} - \hbar\omega},\label{reactKubo} \\
    \mathscr{A}\big(\chi^{i\ell}(\omega)\big) &= \pi e^2 \hbar^2 \sum_{mn} f_{nm} \int_{\mathrm{BZ}^3} \frac{d\bm{k}}{(2\pi)^3}\frac{1}{(E_{m\bm{k}} - E_{n\bm{k}})^2} v_{nm}^i(\bm{k}) v_{mn}^{\ell}(\bm{k})\delta\big(E_{m\bm{k}} - E_{n\bm{k}} - \hbar\omega\big),
    \label{absKubo}
\end{align}
\end{widetext}
Then the reactive and absorptive parts of the effective dielectric tensor are given by
\begin{align}
    \mathscr{R}\big(\varepsilon^{i\ell}(\omega)\big) &= \delta_{i\ell} + 4\pi \mathscr{R}\big(\chi^{i\ell}(\omega)\big) + \frac{2i e^2}{\hbar\omega} \varepsilon^{ij\ell}\, \hat{\bm{e}}_j \cdot \bm{C}_{\mathcal{V}},\nonumber \\
    \mathscr{A}\big(\varepsilon^{i\ell}(\omega)\big) &= 4\pi \mathscr{A}\big(\chi^{i\ell}(\omega)\big).
    \label{reactabsdielectricCI}
\end{align}
The prefactor $f_{nm} = f_n - f_m$ in the absorptive part of the Kubo susceptibility (\ref{absKubo}) is only nonzero if only one of $f_n$ or $f_m$ is nonzero. This means that one of the indices $n,m$ must represent an occupied valence band and the other represents an unoccupied conduction band. Define the quantity $E_{\mathrm{gap}}$ to be the smallest separation between the highest-energy occupied valence band and the lowest-energy unoccupied conduction band across the $\mathrm{BZ}^3$. If we restrict the frequencies $\omega$ of the Maxwell fields to those for which $\hbar \omega < E_{\mathrm{gap}}$, then the absorptive part of the Kubo susceptibility tensor (\ref{absKubo}) must vanishes, even in the absence of time-reversal symmetry.

%%%%%%%%%%%%%%%%%%%%%%%%%%%%%%%%%%%%%%%%%%

\section{Discussion}\label{Sec:Discussion}

We have introduced a second-quantized continuum model for bulk Chern insulators wherein the Hamiltonian (\ref{H0density}) involves a static vector potential that has the periodicity of the lattice and characterizes the spontaneously broken time-reversal symmetry in the system's unperturbed ground state. The static magnetic field it generates can be thought of as arising from one or more ions with intrinsic magnetic moments in each unit cell of the crystal, which holds for experimental realizations of Chern insulators such as magnetically doped $\mathrm{Bi}_2\mathrm{Te}_3$ and $\mathrm{Sb}_2\mathrm{Te}_3$ \cite{QAHExperiment2, QAHExperiment3, QAHExperiment5}, or thin films of the magnetic material $\mathrm{Mn}\mathrm{Bi}_2\mathrm{Te}_4$ \cite{QAHExperiment6} with an odd number of septuple layers. Because the Hamiltonian density (\ref{H0density}) is cell-periodic, we were able to use basic results from condensed matter physics such as Bloch's theorem \cite{Bloch} to analyze the electronic structure of the system. We derived global expressions for the Chern invariants characterizing the topology of Chern insulators -- that is, the Chern number in two dimensions and the Chern vector in three dimensions -- which were written in terms of the velocity matrix elements in the basis of Bloch functions\textcolor{blue}{,} and explicitly featured the static vector potential responsible for the breaking of time-reversal symmetry. 

Our approach is quite versatile: For any material that is found to support a Chern insulator phase, once the crystal's lattice structure $(\mathscr{C},\Gamma)$ and the electrostatic and static vector potentials in the Hamiltonian (\ref{H0density}) are specified, then the Bloch energy eigenfunctions and their cell-periodic parts can be calculated using numerical methods, and the Chern invariants and response tensors can be calculated from our global expressions. There is a fundamental complexity in numerical calculations of these topological invariants from their definitions in terms of the Berry connection and its Berry curvature, since the Chern invariants obtained from these definitions involve complicated Brillouin-zone integrals for which band crossings lead to numerical issues \cite{BZintegration}. These problems in numerical integration are circumvented in our global expressions (\ref{microscopicChernnumberV}) and (\ref{microscopicChernvector}) by the presence of the prefactors $f_{nm} = f_n - f_m$ that vanish at band crossings.

The response of a bulk Chern insulator to applied electric fields at finite frequency was discussed in Sec. \ref{Sec:LinearResponse}, using an expression for the conductivity tensor in the long-wavelength approximation that was introduced in recent work \cite{Kattan2}. There we derived the conductivity tensor from the Hamiltonian (\ref{H0}) using a formalism based on microscopic polarization and magnetization fields in extended media \cite{Mahon2019}, which has also been used to derive macroscopic notions of polarization and magnetization in a Chern insulator analogous to those that feature in the ``modern theories of polarization and magnetization" \cite{CImanuscript1}. We introduced an effective dielectric tensor (\ref{effectivedielectrictensor}) that can be used to study the optical response of Chern insulators, such as the transmission and reflection of light across a thin film or finite-sized sample, or even to focused beams. There are many interesting consequences of the properties of this effective dielectric tensor. For example, the asymmetry of this tensor under index exchange leads to two distinct refractive indices for light propagating in the system, meaning that a Chern insulator is ``optically active." The optical properties of Chern insulators that arise from the effective dielectric tensor (\ref{effectivedielectrictensor}) is the topic of a forthcoming publication.

An advantage of this approach, which is based on a formalism of microscopic polarization and magnetization fields in extended media \cite{Mahon2019} from which the conductivity tensor was derived \cite{Kattan2}, is that we can move beyond the long-wavelength approximation and study the higher-order response of Chern insulators to electric and magnetic fields. For example, there is a well-defined expansion of the macroscopic current density \cite{Mahon2020}
\begin{align}
    J^{i}(\bm{q},\omega) = \sigma^{i\ell}(\omega) E^{\ell}(\bm{q},\omega) + \sigma^{i\ell j}(\omega) E^{\ell}(\bm{q},\omega) q^j + O(q^2),
    \label{qexpansion}
\end{align}
where $\bm{q}$ is the wavevector of the electric field. We have only studied the long-wavelength response associated in the first term in this communication, but it is straightforward to derive higher-order terms in this expansion, such as the second ``$O(q)$ term" accounting for the magnetoelectric effect and optical activity \cite{Mahon2020a}, or the $O(q^2)$ term which includes the magnetic susceptibility \cite{Duff2}. Moreover, it is possible to move beyond linear response and derive bulk non-linear response tensors (\textit{e.g.} the second-order electric susceptibility) for Chern insulators. Higher-order contributions to the expansion (\ref{qexpansion}) and non-linear response are topics of our ongoing work.

%%%%%%%%%%%%%%%%%%%%%%%%%%%%%%%%%%%%%%%%%%

\begin{acknowledgments}
    We thank Alistair H. Duff for helpful discussions. This work was supported by the Natural Sciences and Engineering Research Council of Canada (NSERC).
\end{acknowledgments}

%%%%%%%%%%%%%%%%%%%%%%%%%%%%%%%%%%%%%%%%%%

\appendix

%%%%%%%%%%%%%%%%%%%%%%%%%%%%%%%%%%%%%%%%%%

\section{Quasi-Bloch frames and exponentially localized Wannier functions}\label{Appendix:QuasiBlochframes}

One of the primary advantages of this field theoretic formalism introduced in Sec. \ref{Sec:HamiltonianTheory} is we can study the electronic structure of Chern insulators and their response to electromagnetic fields in terms of such fundamental quantities as polarization, magnetization, and free charges and currents, all of which can be formulated \textit{microscopic level}. This formalism based on microscopic polarization and magnetization fields, and microscopic free charge and current densities, was used in previous work to study the ground-state properties \cite{CImanuscript1} of a Chern insulator, and was used to derive the long-wavelength conductivity tensors (\ref{totalconductivity2D}) and (\ref{totalconductivity3D}) in linear response \cite{Kattan2}. And, in principle, it can be used to study both linear and nonlinear response to arbitrary order in the $\bm{q}$-expansion (\ref{qexpansion}) of the macroscopic current density \cite{Mahon2019, Mahon2020, Mahon2020a}.

These microscopic polarization and magnetization fields are defined in terms of generalized electric and magnetic multipole series expansions, where the multipoles are the moments of the microscopic charge-current distribution localized around each of the lattice sites of the crystal. For these series expansions to be well-defined and convergent, a necessary and sufficient condition is the microscopic charge and current densities be constructed in terms of a complete set of exponentially localized Wannier functions (ELWFs) for the Chern insulator. To define such a set of ELWFs, we first define a collection of \textit{quasi-Bloch functions} through local unitary transformations of the Bloch functions $u_{n\bm{k}}(\bm{x})$ of the form
\begin{align}
    u_{\alpha\bm{k}}(\bm{x}) = \sum_n U_{n\alpha}(\bm{k}) u_{n\bm{k}}(\bm{x}),
    \label{unitarytransformation}
\end{align}
where the sum is over all of the bands and the index $\alpha \in \mathbb{N}$ is generally distinct from the band index $n \in \mathbb{N}$. We can introduce a set of \textit{quasi-Bloch states} $\ket{\alpha\bm{k}}$ such that the corresponding wavefunctions $u_{\alpha\bm{k}}(\bm{x}) = \bra{\bm{x}}\ket{\alpha\bm{k}}$ are the quasi-Bloch functions above. The quantities $U_{n\alpha}(\bm{k})$ are the matrix components of a $\bm{k}$-dependent unitary operator $U(\bm{k})$ acting on the Hilbert space (\ref{BlochHilbertspace}). And for an appropriate choice of unitary operator at each $\bm{k} \in \mathrm{BZ}^d$ it is always possible to construct these quasi-Bloch states in such a way that the mapping $\bm{k} \mapsto (\ket{\alpha\bm{k}})_{\alpha = 1}^{\infty}$ is globally smooth across the $\mathrm{BZ}^d$, including at points in the locus of degeneracies $\mathscr{D}_{\mathcal{B}}$ \cite{CImanuscript1}. These globally smooth maps will be called \textit{quasi-Bloch frames}, and they can be used to define a set of exponentially localized Wannier functions
\begin{align}
    W_{\alpha\bm{R}}(\bm{x}) = \sqrt{\Omega_{\mathrm{uc}}} \int_{\mathrm{BZ}^d}\frac{d\bm{k}}{(2\pi)^d} e^{i\bm{k}\cdot(\bm{x}-\bm{R})} u_{\alpha\bm{k}}(\bm{x})
    \label{ELWF}
\end{align}
for each lattice site $\bm{R} \in \Gamma$. It can be shown \cite{Brouder, Kattan1} that the exponential localization of these Wannier functions is equivalent to the quasi-Bloch functions $u_{\alpha\bm{k}}(\bm{x})$ being holomorphic in a tubular region of the complex space $\mathbb{C}^d$ containing the Brillouin zone. Provided all of the bands are included in the sum (\ref{unitarytransformation}), such quasi-Bloch functions can always be found, and thus it is always possible to form ELWFs for a Chern insulator, albeit ones that include at least some unoccupied conduction bands in additional to the occupied valence bands in their definition. 

A common theme of our discussion in Sec. \ref{Sec:ChernInvariants} was the local definition of quantities such as the Berry connection (\ref{BerryconnectionBcomponents}) and its Berry curvature (\ref{BerrycurvatureBcomponents}), for example, and their projected counterparts on the valence bundle. An advantage of this approach based on quasi-Bloch states is that we can define gauge potentials for the Berry connection $1$-form and the Berry curvature $2$-form that are \textit{globally smooth} across the Brillouin zone. For example, the Berry connection $1$-form can be expanded in terms of such a global quasi-Bloch frame as
\begin{align}
    \xi(\bm{k}) = \sum_{\alpha\beta} \tilde{\xi}_{\alpha\beta}(\bm{k}) \ketbra{\alpha\bm{k}},
    \label{BerryconnectionBtilde}
\end{align}
where the sum includes all $\alpha,\beta \in \mathbb{N}$ and the corresponding gauge potential is given by \cite{Kattan2}
\begin{equation}
    \tilde{\xi}_{\alpha\beta}(\bm{k}) = i (\alpha\bm{k}|\mathrm{d} \beta\bm{k}),
    \label{BerryconnectionBcomponentstilde}
\end{equation}
involving the exterior derivative $\mathrm{d} = \mathrm{d}k^a \partial / \partial k^a$ and the inner product (\ref{innerproduct}). Unlike the gauge potential (\ref{BerryconnectionBcomponents}) in a local Bloch frame, this gauge potential in a quasi-Bloch frame is globally smooth across the Brillouin zone. Working in a Cartesian chart for the $\mathrm{BZ}^d$, the components of the non-Abelian Berry connection in this quasi-Bloch frame are \cite{VanderbiltBook}
\begin{align}
    \tilde{\xi}_{\alpha\beta}^i(\bm{k}) = \frac{i}{\Omega_{\mathrm{uc}}} \int_{\Omega} d\bm{x}\, u_{\alpha\bm{k}}^{\dagger}(\bm{x}) \frac{\partial u_{\beta\bm{k}}(\bm{x})}{\partial k^i}.
    \label{Wannierconnectioncomponents}
\end{align}
Recalling that the Bloch functions and these quasi-Bloch functions are related by Eq. (\ref{unitarytransformation}), the gauge potentials (\ref{BerryconnectionBcomponents}) and (\ref{BerryconnectionBcomponentstilde}) for the Berry connection can be shown to be related by the unitary transformation rule
\begin{align}
    \sum_{\alpha\beta} U_{m\alpha}(\bm{k}) \tilde{\xi}_{\alpha\beta}^i(\bm{k}) U_{\beta n}^{\dagger}(\bm{k}) = \xi_{mn}^i(\bm{k}) + \mathcal{W}_{mn}^i(\bm{k}),
    \label{BlochWannierconnection}
\end{align}
featuring the Bloch-frame matrix elements
\begin{align}
    \mathcal{W}_{mn}^i(\bm{k}) \equiv i \sum_{\alpha} \frac{\partial U_{m\alpha}(\bm{k})}{\partial k^i} U_{\alpha n}^{\dagger}(\bm{k}).
    \label{Wmatrixcomponents}
\end{align}
As is the case for all Bloch-frame quantities, the transformation rule (\ref{BlochWannierconnection}) and these matrix elements are only defined away from the local of degeneracies, even though the quasi-Bloch frame components (\ref{Wannierconnectioncomponents}) are globally smooth. Implementing the expansion (\ref{BerryconnectionBtilde}) in the definition (\ref{BerrycurvatureB}) of the  curvature $2$-form, the Berry curvature of the globally smooth gauge potential (\ref{BerryconnectionBcomponentstilde}) in a quasi-Bloch frame is the globally smooth, matrix-valued $2$-form
\begin{align}
    \tilde{F}_{\alpha\beta}(\bm{k}) = \mathrm{d}\tilde{\xi}_{\alpha\beta}(\bm{k}) - i \sum_{\gamma} \tilde{\xi}_{\alpha\gamma}(\bm{k}) \wedge \tilde{\xi}_{\gamma\beta}(\bm{k}),
    \label{BerrycurvatureBcomponentstilde}
\end{align}
and the analogue of the local Bloch-frame identity (\ref{CurvaturefreeBlochframe}) in a quasi-Bloch frame is the global identity 
\begin{align}
    \varepsilon^{iab} \partial_a \tilde{\xi}_{\alpha\beta}^b(\bm{k}) &= i\sum_{\gamma} \varepsilon^{iab} \tilde{\xi}_{\alpha\gamma}^a(\bm{k}) \tilde{\xi}_{\gamma\beta}^b(\bm{k}),
    \label{CurvaturefreeWannierframe}
\end{align}
valid everywhere in the Brillouin zone. However, unlike the curvature $2$-form (\ref{BerrycurvatureB}) for the Berry connection on the Bloch bundle, the curvature $2$-form for the corresponding projected connection on the valence bundle does not admit any expression analogous to Eq. (\ref{BerrycurvatureV}) in a quasi-Bloch frame, except in the case of a topologically trivial insulator. This is related to the fact that one cannot define an ``occupation factor" $f_{\alpha}$ for the quasi-Bloch functions (\ref{unitarytransformation}) in a Chern insulator, since to form such wavefunctions that are globally smooth across the Brillouin zone, we need to include the valence bands \textit{and} a sufficient number of conduction bands in the sum so that the Chern invariant for all of these bands together vanishes. This is unlike the scenario for a topologically trivial insulator. There the sum in Eq. (\ref{unitarytransformation}) can be restricted to the valence bands alone, and then we can simply define $f_{\alpha} = f_{n}$ in terms of the Fermi occupation factor for the Bloch functions.

%%%%%%%%%%%%%%%%%%%%%%%%%%%%%%%%%%%%%%%%%%

\section{Symmetries in the tenfold way}\label{Appendix:Symmetries}

We study the symmetry properties of our field theory introduced in Sec. \ref{Sec:HamiltonianTheory}, and the global expressions (\ref{microscopicChernnumberV}) and (\ref{microscopicChernvector}) for the Chern invariants, in the context of the three discrete symmetries of the tenfold way classification scheme. The implementation of discrete symmetries on the electronic Fock space $\mathcal{H}_F$ is formulated within the representation theory of finite groups \cite{NakaharaBook}. Consider a group $G$ of symmetry transformations that are linearly realized on the single-particle Hilbert space $\mathcal{H}_1$ through a unitary (or antiunitary) representation $\mathcal{U} : G \to \mathrm{U}(\mathcal{H}_1)$ that assigns to each group element $g \in G$ a unitary (antiunitary) operator $\mathcal{U}_g \in \mathrm{U}(\mathcal{H}_1)$. This representation lifts to a unitary (antiunitary) representation $U : G \to \mathrm{U}(\mathcal{H}_F)$ on the Fock space $\mathcal{H}_F$ assigning to each group element $g \in G$ a unitary (antiunitary) operator $\hat{U}_g$ acting on this Fock space. We say that the group $G$ is a \textit{symmetry} if and only if the following condition holds:
\begin{align}
    \big[\hat{U}_g, \hat{H}_0\big]_- = 0 \;\; \iff \;\; \hat{U}_g \hat{H}_0 \hat{U}_g^{-1} = \hat{H}_0
\end{align}
for all $g \in G$, involving the second-quantized Hamiltonian $\hat{H}_0$ in the Schrödinger representation. Importantly, since the unitary (anti-unitary) operators $\hat{U}_g$ act on the \textit{electronic} Fock space, and the associated unitary (antiunitary) operators $\mathcal{U}_g$ act on the single-particle \textit{electronic} Hilbert space, these symmetry operations are effected only on the electronic degrees of freedom, and \textit{not} on the electrostatic potential $\mathrm{V}_{\Gamma}(\bm{x})$ or the static vector potential $\bm{a}_{\mathrm{static}}(\bm{x})$ in the Hamiltonian density (\ref{H0density}), both of which are classical fields.

\subsection{Time-reversal symmetry}\label{Appendix:TimeReversal}

Time-reversal transformations are represented by an antiunitary operator $\hat{T}$ acting on the electronic Fock space, and we denote by $\mathcal{T}$ the corresponding time-reversal operator on the single-particle Hilbert space. The time-reversal transformation of the electron field operator is given by \cite{Ryu2010}
\begin{align}
    \hat{T} \hat{\psi}_{\sigma}(\bm{x}) \hat{T}^{-1} = (U_T)_{\sigma\;\;}^{\;\;\sigma'} \hat{\psi}_{\sigma'}(\bm{x}),
\end{align}
where $U_T$ is a unitary operator acting on the spin sector. We say that the second-quantized Hamiltonian (\ref{H0}) is \textit{time-reversal symmetric} if and only if the following condition holds:
\begin{align}
    \mathcal{T}^{\dagger} \mathcal{H}_0(\bm{x}) \mathcal{T} = \mathcal{H}_0(\bm{x}),
\end{align}
where we have defined $\mathcal{T} \equiv U_T \mathcal{K}$ with $\mathcal{K}$ being the complex-conjugation operator. It is well-known that for spin-$1/2$ electrons the unitary matrix $U_T = i \sigma_2$ leading to $\mathcal{T} = i\sigma_2 \mathcal{K}$, where $\sigma_2$ is the Pauli $y$-matrix \cite{Ryu2010}. It is straightforward to show that the Hamiltonian density (\ref{H0density}) transforms as
\begin{align}
    \mathcal{T}^{\dagger} \mathcal{H}_0(\bm{x})\mathcal{T} = \frac{1}{2m}\big(\bm{\mathfrak{p}}^*(\bm{x})\big)^2 + \mathrm{V}_{\Gamma}(\bm{x}) + \frac{e\hbar}{2mc} \bm{\sigma}\cdot\bm{b}_{\mathrm{static}}(\bm{x})\nonumber \\
    - \frac{\hbar}{4m^2 c^2} \bm{\sigma}\cdot\bm{\nabla}\mathrm{V}_{\Gamma}(\bm{x}) \times \bm{\mathfrak{p}}^*(\bm{x}),
\end{align}
which is \textit{not} equal to Eq. (\ref{H0density}) and so time-reversal symmetry is broken, as expected. However, suppose that the static vector potential and magnetic field in this Hamiltonian (\ref{H0density}) vanishes so that $\bm{\mathfrak{p}}(\bm{x}) \to \bm{p}(\bm{x}) = (\hbar/i)\bm{\nabla}$. Then it is straightforward to show that time-reversal symmetry holds, and this time-reversal invariant Hamiltonian density is given by
\begin{align}
    \mathcal{H}_0(\bm{x}) &\overset{\mathcal{T}}{=} \frac{1}{2m}\big(\bm{p}(\bm{x})\big)^2 + \mathrm{V}_{\Gamma}(\bm{x})\nonumber \\
    &+ \frac{\hbar}{4m^2 c^2} \bm{\sigma}\cdot\bm{\nabla}\mathrm{V}_{\Gamma}(\bm{x}) \times \bm{p}(\bm{x}),
\end{align}
where the Zeeman term vanishes and the third line describes spin-orbit coupling in the absence of the static vector potential.

When time-reversal symmetry holds, it should follow that the Chern invariants in both two and three dimensions must vanish. We first observe that under time-reversal transformations the Bloch functions satisfy \cite{Duff1} 
\begin{align}
    \mathcal{T} u_{n\bm{k}}(\bm{x}) \overset{\mathcal{T}}{=} e^{-i\lambda_n(\bm{k})} u_{n(-\bm{k})}(\bm{x}),
    \label{BlochwavefunctionsTRS}    
\end{align}
where $\lambda_n(\bm{k})$ is a phase and the notation ``$\overset{\mathcal{T}}{=}$" indicates \textcolor{blue}{as above} an equality that holds only in the presence of time-reversal symmetry. This leads to a transformation of the velocity matrix elements
\begin{align}
    v_{nm}^i(\bm{k}) \overset{\mathcal{T}}{=}  e^{i(\lambda_m(\bm{k}) - \lambda_n(\bm{k}))} v_{mn}^i(-\bm{k}),
\end{align}
in which case these global expressions pick up a minus sign upon relabelling of the summation indices. And then, since the energies $E_{n\bm{k}}$ and $E_{n(-\bm{k})}$ are equal when time-reversal symmetry holds, the integrand in Eqs. (\ref{microscopicChernnumberV}) and (\ref{microscopicChernvector}) is odd and thereby vanishes upon integration. Therefore the Chern number $C_{\mathcal{V}}$ in two dimensions vanishes, as does the Chern vector $\bm{C}_{\mathcal{V}}$ in three dimensions. It should be emphasized, however, that the converse is not true; the Chern invariants may still vanish when time-reversal symmetry is broken, such as in a topologically trivial ferromagnet.

\subsection{Particle-hole symmetry}\label{Appendix:ParticleHole}

Particle-hole transformations are represented by a \textit{unitary} operator $\hat{C}$ acting on the electronic Fock space, and we denote by $\mathcal{C}$ the corresponding particle-hole operator on the single-particle Hilbert space. The particle-hole transformation of the electron field operator is
\begin{align}
    \hat{C} \hat{\psi}_{\sigma}(\bm{x}) \hat{C}^{-1} = (U_C^*)_{\sigma\;\;}^{\;\;\sigma'} \hat{\psi}_{\sigma'}^{\dagger}(\bm{x}),
\end{align}
where $U_C$ is a unitary operator acting on the spin sector. We say that the second-quantized Hamiltonian (\ref{H0}) is \textit{particle-hole symmetric} if and only if the following condition holds:
\begin{align}
    \mathcal{C}^{\dagger} \mathcal{H}_0(\bm{x}) \mathcal{C} = -\mathcal{H}_0(\bm{x}) \;\; \text{and} \;\; \Tr \mathcal{H}_0(\bm{x}) = 0,
\end{align}
where we have defined $\mathcal{C} \equiv U_C \mathcal{K}$. It is straightforward to show that these conditions on the Hamiltonian density (\ref{H0density}) never hold, since the trace condition never holds for the first two terms of this Hamiltonian density (\ref{H0density}). However, this condition does hold for many superconducting systems where $U_C = \sigma_1$ so that $\mathcal{C} = \sigma_1 \mathcal{K}$. Therefore our system has broken particle-hole symmetry, which is to be expected as this symmetry is often only respected by superconductors and chiral topological insulators \cite{Ryu2010}.

\subsection{Sublattice symmetry}\label{Appendix:Chiral}

Sublattice transformations are represented by an antiunitary operator $\hat{S} = \hat{T} \circ \hat{C}$ acting on the electronic Fock space that is the composition of the particle-hole and time-reversal operators, and we denote by $\mathcal{S}$ the corresponding sublattice operator on the single-particle Hilbert space. The sublattice transformation of the electron field operator is 
\begin{align}
    \hat{S} \hat{\psi}_{\sigma}(\bm{x}) \hat{S}^{-1} = (U_S^*)_{\sigma\;\;}^{\;\;\sigma'} \hat{\psi}_{\sigma'}^{\dagger}(\bm{x}),
\end{align}
where $U_S = U_C^* U_T^*$ is a unitary operator acting on the spin sector. We say that the second-quantized Hamiltonian (\ref{H0}) is \textit{sublattice symmetric} if and only if the following condition holds:
\begin{align}
    \mathcal{S}^{\dagger} \mathcal{H}_0(\bm{x}) \mathcal{S} = -\mathcal{H}_0(\bm{x}) \;\; \text{and} \;\; \Tr \mathcal{H}_0(\bm{x}) = 0,
\end{align}
where we have defined the sublattice operator $\mathcal{S} \equiv \mathcal{T} \circ \mathcal{C}$. As in the case of particle-hole symmetry, it is straightforward to show that these conditions on the Hamiltonian density (\ref{H0density}) never hold. Notably, for superconducting systems with $\mathcal{C} = \sigma_1 \mathcal{K}$ we have $\mathcal{S} = \sigma_3$.

\section{Gauge invariance}\label{Ch.3.Sec:Gaugetransformations}

The global expressions (\ref{microscopicChernnumberV}) and (\ref{microscopicChernvector}) for the Chern invariants in two and three dimensions feature the matrix elements (\ref{velocitymatrixelements}) of the velocity operator (\ref{velocityoperator}), which involves the static vector potential instead of the magnetic field (\ref{staticB}) it generates. This vector potential is subject to the $\mathrm{U}(1)$--gauge transformations
\begin{align}
    \bm{a}_{\mathrm{static}}'(\bm{x}) = \bm{a}_{\mathrm{static}}(\bm{x}) + \bm{\nabla}g(\bm{x}),
    \label{gaugetransformation}
\end{align}
where $g(\bm{x})$ is a gauge function. Since we work in the frozen-ion approximation, it follows that the generating function must be cell-periodic, with $g(\bm{x} + \bm{R}_{\mathrm{t}}) = g(\bm{x})$ for all lattice translations $\bm{R}_{\mathrm{t}} \in \Gamma$. For the global expressions (\ref{microscopicChernnumberV}) and (\ref{microscopicChernvector}) to be well-defined, we need to show the invariance of these expressions under such $\mathrm{U}(1)$--gauge transformations of the static vector potential. To do so, it is useful to rewrite the velocity matrix elements (\ref{velocitymatrixelements}) in terms of an appropriate set of exponentially localized Wannier functions (\ref{ELWF}). In the presence of a nonuniform vector potential, it is possible to find a particular such set of Wannier functions that can be written as \cite{Duff1}
\begin{align}
    W_{n\bm{R}}(\bm{x}) = e^{i\Phi^{s}(\bm{x},\bm{R})} \chi_{n\bm{R}}^{s}(\bm{x}),
    \label{modWF}
\end{align}
where the superscript ``$s$" indicates that only the static quantities in Eq. (\ref{staticB}) are involved, and the wavefunctions $\chi_{n\bm{R}}^{s}(\bm{x})$ on the right-hand-side depend only on the magnetic field (\ref{staticB}) and are therefore gauge-invariant. Here we have introduced a generalized Peierls phase 
\begin{align}
    \Phi^{s}(\bm{x},\bm{R}) = \frac{e}{\hbar c} \int d\bm{y}\, s^i(\bm{y};\bm{x},\bm{R}) a_{\mathrm{static}}^i(\bm{y}),
    \label{generalizedPeierlsphase}
\end{align}
involving a ``relator" \cite{Mahon2019}
\begin{align}
    \bm{s}(\bm{y};\bm{x},\bm{R}) = \int_{C(\bm{x},\bm{R})} d\bm{z}\, \delta(\bm{z} - \bm{y}),
\end{align}
where $\bm{z} : \mathbb{R} \to \mathbb{R}^d$ is a smooth curve with image path $C(\bm{x},\bm{R})$ that begins at a lattice site $\bm{R} \in \Gamma$ and ends at a point $\bm{x} \in \mathbb{R}^d$. From the line integral (\ref{generalizedPeierlsphase}) of the static vector potential, we can also define
\begin{align}
    \Delta^{s}(\bm{x},\bm{y},\bm{z}) \equiv \Phi^{s}(\bm{x},\bm{y}) + \Phi^{s}(\bm{y},\bm{z}) + \Phi^{s}(\bm{z},\bm{x}),
\end{align}
which is the flux of the magnetic field (\ref{staticB}) through the surface that is bounded by the curves $C(\bm{x},\bm{y})$, $C(\bm{y},\bm{z})$, and $C(\bm{z},\bm{x})$. Then the velocity matrix elements (\ref{velocitymatrixelements}) can be written in terms of these Wannier functions as
\begin{align}
    \bm{v}_{nm}(\bm{k}) &= \frac{\Omega_{\mathrm{uc}}}{(2\pi)^d} \sum_{\bm{R}_1\bm{R}_2} e^{i\bm{k}\cdot(\bm{R}_2 - \bm{R}_1)} e^{i\Phi^{s}(\bm{R}_1,\bm{R}_2)}\nonumber \\
    &\times \int d\bm{x}\, e^{i\Delta^{s}(\bm{R}_1,\bm{x},\bm{R}_2)} \chi_{n\bm{R}_1}^{s\dagger}(\bm{x}) \bm{v}(\bm{x},\bm{R}_2) \chi_{m\bm{R}_2}^{s}(\bm{x}),
    \label{frakturpWF}
\end{align}
where we have defined the modified velocity operator
\begin{align}
    \bm{v}(\bm{x},\bm{R}) \equiv \frac{1}{m} \left(\frac{\hbar}{i} \bm{\nabla} - \frac{e}{c} \bm{\Omega}_{\bm{R}}^s(\bm{x})\right) + \frac{\hbar}{4m^2 c^2} \bm{\sigma}\times \bm{\nabla}\mathrm{V}_{\Gamma}(\bm{x})
\end{align}
which features the static magnetic field (\ref{staticB}) through 
\begin{align}
    \Omega_{\bm{R}}^{s,i}(\bm{x}) = \int d\bm{y}\, \alpha^{ji}(\bm{y};\bm{x},\bm{R}) b_{\mathrm{static}}^j(\bm{y}),
\end{align}
and we have introduced another ``relator" \cite{Mahon2019}
\begin{align}
    \alpha^{ji}(\bm{y};\bm{x},\bm{R}) = \varepsilon^{jmn} \int_{C(\bm{x},\bm{R})} dz^m \frac{\partial z^n}{\partial x^i} \delta(\bm{z} - \bm{y}).
\end{align}
The only term in Eq. (\ref{frakturpWF}) that is not manifestly gauge-invariant is the generalized Peierls phase, which under the gauge transformation (\ref{gaugetransformation}) becomes
\begin{align}
    \Phi^{s}{}'(\bm{R}_1,\bm{R}_2) = \Phi^s(\bm{R}_1,\bm{R}_2) + \frac{e}{\hbar c}\Big(g(\bm{R}_1) - g(\bm{R}_2)\Big).
\end{align}
But the second term vanishes because the gauge function $g(\bm{x})$ is cell-periodic, and so these velocity matrix elements are indeed gauge-invariant. We can also write the energies in the denominator of Eq. (\ref{microscopicChernnumberV3D}) as
\begin{align}
    E_{n\bm{k}} &= \frac{\Omega_{\mathrm{uc}}}{(2\pi)^d} \sum_{\bm{R}_1\bm{R}_2} e^{i\bm{k}\cdot(\bm{R}_2-\bm{R}_1)} e^{i\Phi^s(\bm{R}_1,\bm{R}_2)}\nonumber \\
    &\times \int d\bm{x}\, e^{i\Delta^s(\bm{R}_1,\bm{x},\bm{R}_2)} \chi_{n\bm{R}_1}^{s\dagger}(\bm{x}) \mathcal{H}_0(\bm{x},\bm{R}_2) \chi_{n\bm{R}_2}^s(\bm{x}),
\end{align}
where the gauge-invariant Hamiltonian density is
\begin{align}
    \mathcal{H}_0(\bm{x},\bm{R}) &= \frac{1}{2m}\left(\frac{\hbar}{i}\bm{\nabla} - \frac{e}{c} \bm{\Omega}_{\bm{R}}^s(\bm{x})\right)^2 + \mathrm{V}_{\Gamma}(\bm{x})\nonumber \\
    &+ \frac{\hbar}{4m^2 c^2} \bm{\sigma}\cdot\bm{\nabla}\mathrm{V}_{\Gamma}(\bm{x}) \times \left(\frac{\hbar}{i}\bm{\nabla} - \frac{e}{c} \bm{\Omega}_{\bm{R}}^s(\bm{x})\right)\nonumber \\
    &- \frac{e\hbar}{2mc} \bm{\sigma}\cdot\bm{b}_{\mathrm{static}}(\bm{x})
\end{align}
Since all of the quantities in this expression are gauge-invariant, it follows that our global expressions (\ref{microscopicChernnumberV}) and (\ref{microscopicChernvector}) for the Chern invariants are indeed gauge-invariant.

%%%%%%%%%%%%%%%%%%%%%%%%%%%%%%%%%%%%%%%%%%

\section{Kubo-Greenwood formula}\label{Appendix:KuboGreenwood}

We show that the Kubo-Greenwood formula (\ref{KuboGreenwood}) reproduces our conductivity tensors (\ref{totalconductivity2D}) and (\ref{totalconductivity3D}) for Chern insulators in two and three dimensions when the $\bm{k}$-dependent Hamiltonian density (\ref{H0densityk}) for our field theory is used in the matrix elements (\ref{matrixpartialH}). We separate the denominators in the integrand of Eq. (\ref{KuboGreenwood}) using
\begin{widetext}
\begin{align}
    \frac{1}{E_{m\bm{k}} - E_{n\bm{k}}}\frac{1}{E_{m\bm{k}} - E_{n\bm{k}} - \hbar(\omega+i0^+)} = \frac{1}{(E_{m\bm{k}} - E_{n\bm{k}})^2} + \frac{\hbar\omega}{(E_{m\bm{k}} - E_{n\bm{k}})^2 (E_{m\bm{k}} - E_{n\bm{k}} - \hbar(\omega+i0^+))},
\end{align}
from which follows
\begin{align}
    \sigma_{\mathrm{KG}}^{i\ell}(\omega) &= \frac{ie^2}{\hbar} \sum_{mn} f_{nm} \int_{\mathrm{BZ}^d}\frac{d\bm{k}}{(2\pi)^d} \frac{1}{(E_{n\bm{k}} - E_{m\bm{k}})(E_{m\bm{k}} - E_{n\bm{k}})} (n\bm{k}|\partial_{i}H_{\bm{k}}|m\bm{k})(m\bm{k}|\partial_{\ell}H_{\bm{k}}|n\bm{k})\nonumber \\
    &+ i\omega e^2 \sum_{mn} f_{nm} \int_{\mathrm{BZ}^d}\frac{d\bm{k}}{(2\pi)^d} \frac{1}{(E_{n\bm{k}} - E_{m\bm{k}})(E_{m\bm{k}} - E_{n\bm{k}})} \frac{(n\bm{k}|\partial_{i}H_{\bm{k}}|m\bm{k})(m\bm{k}|\partial_{\ell}H_{\bm{k}}|n\bm{k})}{E_{m\bm{k}}-E_{n\bm{k}} - \hbar(\omega + i 0^+)},
    \label{KuboGreenwood2}
\end{align}
\end{widetext}
where the matrix elements in the numerator are given by Eq. (\ref{matrixpartialH}). From our $\bm{k}$-dependent Hamiltonian density (\ref{H0densityk}), these matrix elements are given by
\begin{align}
    (n\bm{k}|\partial_{i}H_{\bm{k}}|m\bm{k}) &= \hbar v_{nm}^{i}(\bm{k}) + \frac{\hbar k^i}{2m} \delta_{nm},
\end{align}
involving the velocity matrix elements (\ref{velocitymatrixelements}), and the matrix elements in the integrand above are
\begin{align}
    \sum_{mn} f_{nm} (n\bm{k}|\partial_{i}H_{\bm{k}}|m\bm{k})(m\bm{k}|\partial_{\ell}H_{\bm{k}}|n\bm{k})\nonumber \\
    = \hbar^2 \sum_{mn} f_{nm} v_{nm}^{i}(\bm{k})v_{mn}^{\ell}(\bm{k}).
    \label{firstlineidentity}
\end{align}
To write these quantities in terms of the Bloch-frame components (\ref{BerryconnectionBcomponents}) of the non-Abelian Berry connection, we employ the identity Eq. (\ref{velocityBerryconnection}) leading to
\begin{align}
    &\sigma_{\mathrm{KG}}^{i\ell}(\omega) = - \frac{ie^2}{\hbar} \sum_{mn} f_{nm} \int_{\mathrm{BZ}^d}\frac{d\bm{k}}{(2\pi)^d}\xi_{nm}^{i}(\bm{k})\xi_{mn}^{\ell}(\bm{k})\nonumber \\
    &- i\omega e^2 \sum_{mn} f_{nm} \int_{\mathrm{BZ}^d}\frac{d\bm{k}}{(2\pi)^d} \frac{\xi_{nm}^{i}(\bm{k})\xi_{mn}^{\ell}(\bm{k})}{E_{m\bm{k}}-E_{n\bm{k}} - \hbar(\omega + i 0^+)},
    \label{KuboGreenwood3}
\end{align}
where we recognize the second line as our Kubo conductivity tensor (\ref{Kuboconductivity}). In $d = 2$ dimensions, we simplify the first line using the two-dimensional analogue of the identity (\ref{CurvaturefreeBlochframe}), along with the chart representation (\ref{ChernnumberV2}) of the Chern number, from which follows
\begin{align}
    \sigma_{\mathrm{KG}}^{i\ell}(\omega) = \sigma_{\mathrm{K}}^{i\ell}(\omega) - \frac{e^2}{2\pi \hbar} \varepsilon^{i\ell} C_{\mathcal{V}},
\end{align}
in agreement with our conductivity tensor (\ref{totalconductivity2D}) in two dimensions. Meanwhile, to simplify Eq. (\ref{KuboGreenwood3}) in $d = 3$ dimensions we use the identity (\ref{CurvaturefreeBlochframe}) to rewrite the first line in terms of the diagonal components of the non-Abelian Berry connection, from which follows
\begin{align}
    \sigma_{\mathrm{KG}}^{i\ell}(\omega) = \sigma_{\mathrm{K}}^{i\ell}(\omega) + \frac{e^2}{\hbar} \varepsilon^{i j \ell} \sum_{n} f_n \int_{\mathrm{BZ}^3}\frac{d\bm{k}}{(2\pi)^3} \varepsilon^{jab} \partial_a \xi_{nn}^b(\bm{k}).
\end{align}
But we have shown in previous work \cite{Kattan2} that the integral in the second term can be written
\begin{align}
    \frac{1}{4\pi^2} \sum_{n} f_n \int_{\mathrm{BZ}^3} d\bm{k}\, \varepsilon^{jab} \partial_a \xi_{nn}^b(\bm{k}) = \sum_{\alpha = 1}^3 \big(\hat{\bm{e}}_j \cdot \bm{g}_{\alpha}\big) C_{\mathcal{V}}^{\alpha},
    \label{integralidentityCv}
\end{align}
involving the first Chern numbers (\ref{Chernnumbers3DV}) for the valence bundle over the three fundamental $2$-cycles in the three-dimensional Brillouin zone $\mathrm{BZ}^3$, and so we have
\begin{align}
    \sigma_{\mathrm{KG}}^{i\ell}(\omega) &= \sigma_{\mathrm{K}}^{i\ell}(\omega) - \frac{e^2}{2\pi \hbar} \varepsilon^{i \ell j} \hat{\bm{e}}_j \cdot \bm{C}_{\mathcal{V}},
\end{align}
where in the second term we have used the definition of the Chern vector (\ref{Chernvector}), and so the Kubo-Greenwood formula reproduces our conductivity tensor (\ref{totalconductivity3D}) in three dimensions.

%%%%%%%%%%%%%%%%%%%%%%%%%%%%%%%%%%%%%%%%%%

\bibliographystyle{apsrev4-2}
\bibliography{ChernInsulator.bib}

\end{document}